\documentstyle[sprocl]{article}
\def\be{\begin{equation}}
\def\ee{\end{equation}}
\def\bea{\begin{eqnarray}}
\def\eea{\end{eqnarray}}

\input epsf.tex

\begin{document}
\title{FOUR LECTURES ON M-THEORY\footnote{To appear in proceedings of the 
1996 ICTP Summer School in High Energy Physics and Cosmology (Trieste, June
10-26). hep-th/9612121.}}
\author{P.K. TOWNSEND}
\address{DAMTP, UNIVERSITY OF CAMBRIDGE,\\ 
SILVER ST., CAMBRIDGE, U.K.}
%
%

\maketitle
\abstracts{Synopsis: (i) how superstring theories are unified by M-theory; (ii)
how superstring and supermembrane properties follow from the D=10 and D=11
supersymmetry algebras; (iii) how D=10 and D=11 supergravity theories 
determine the strong coupling limit of superstring theories; (iv) how
properties of Type II p-branes follow from those of M-branes.}

\section{M-theory unification}

Over the decade from 1984-94 superstring theory came to be regarded as the most
promising approach to a unification of all the fundamental forces. The most
compelling argument in its favour is that it naturally describes an ultraviolet
finite and unitary perturbative quantum gravity \cite{berk}. That this is true
only in a spacetime of ten dimensions (D=10) is not so much a problem as an
invitation to find a realistic model of particle physics by compactification to
D=4 \cite{cand}. However, D=10 `superstring theory' is not  a single theory but
actually a collection of five of them. They are \cite{GSW}
\vskip 0.5cm
{\obeylines
(i) Type IIA ,
(ii) Type IIB,
(iii) $E_8\times E_8$ heterotic,
(iv) $SO(32)$ heterotic,
(v) Type I. }
\vskip 0.5cm
\noindent
While the $E_8\times E_8$ heterotic string \cite{gross} is the one favoured in
attempts to make contact with particle physics, the other four are equally
acceptable as perturbative theories of quantum gravity (whereas the bosonic string
is not because of the tachyon in its spectrum). 

In the infinite tension limit each of these five superstring theories is
approximated by its effective field theory, which is an anomaly-free D=10
supergravity theory \cite{agwit,gsanom}. There are also five of these but one of
them, N=1 supergravity/Yang-Mills (YM) theory with gauge group $U(1)^{496}$, is
not the effective field theory of any superstring theory. The other four are as
follows, labelled according to the superstring theory with which each is
associated:
\vskip 0.5cm
{\obeylines
(i) Non-chiral N=2 supergravity (IIA)
(ii) Chiral N=2 supergravity (IIB),
(iii) N=1 supergravity/YM with $E_8\times E_8$ gauge group,
(iv) \& (v) N=1 supergravity/YM with $SO(32)$ gauge group.}
\vskip 0.5cm\noindent
Observe that two of the five superstring theories have equivalent effective
field theories. This `coincidence' led to some early speculation that perhaps
these two superstring theories are really the same theory. If so, the
equivalence must be non-perturbative because the perturbative spectra
are quite different, the heterotic string being a closed oriented string and
the Type I string an unoriented one that may be open or closed.

Unfortunately, or perhaps inevitably, superstring theories are defined only as
divergent asymptotic power series in the string coupling constant $g_s$, which
is related to the expectation value of the massless dilaton field
$\phi$ appearing in the effective supergravity theory, specifically $g_s=
e^{\langle \phi\rangle}$. Thus, superstring theories fail to qualify as truly
unified theories on two counts: (i) there is more than one of them, and (ii)
they are defined only as asymptotic expansions. The essence of recent progress
on the unification front is the realization that by solving problem (ii) we also
solve problem (i), i.e. all five superstring theories are asymptotic expansions
around different vacua of a single non-perturbative theory. This is not to say
that problem (ii) has been solved, but a convincing picture of how the different
asymptotic expansions are unified by a single theory has emerged. The surprising
feature is that this unified theory is actually 11-dimensional at almost all
points in its moduli space of vacua! The well-known fact that D=10 is the
critical dimension  of superstring theory means only that the dimension of
spacetime is {\it at least} ten, because there may be dimensions that are
invisible in perturbation theory. Considerations of supersymmetry imply that the
spacetime dimension cannot exceed eleven, so we are left with either D=10
or D=11 as the actual dimension of spacetime. In this first lecture I shall
attempt to provide an overview of this D=11 superunification and, since the idea
of superunification in D=11 is quite an old one, I shall begin with a brief
sketch of its history.

The idea first arose from the 1978 construction of D=11 
supergravity \cite{cjs} and was closely tied to the revival of the Kaluza-Klein
(KK) program \cite{kksugra}. This program faced insuperable difficulties, e.g.
the non-renormalizability of D=11 supergravity and its failure to admit chiral
compactifications. Both problems were resolved by superstring theory but
the price was a retreat from D=11 to D=10. Only the fact that IIA
supergravity is the dimensional reduction of D=11 supergravity \cite{west}
offered hope of a possible role for D=11. But whereas the 2-form potential of 
D=10 supergravity theories is naturally associated with a string, the 3-form
potential of D=11 supergravity is naturally associated with a membrane
\cite{bst}. While it was known how to incorporate spacetime-supersymmetry into
string theory, via the Green-Schwarz (GS) worldsheet action \cite{greens}, it was
unclear how to generalize this to higher-dimensional objects. Progress came from
consideration of effective actions for extended objects in supersymmetric field
theories; it was known that the D=4 GS action could be
re-interpreted as the effective action for Nielsen Olesen vortices in an N=2
supersymmetric abelian Higgs model, but this model is the dimensional reduction
to D=4 of a D=6 field theory for which the `vortices' are what we would now call
3-branes. The effective action for this D=6 3-brane is necessarily a
higher-dimensional generalization of the  GS action, and its construction 
\cite{hlp} showed how to overcome the `string barrier'. This led to
the construction of the D=11 supermembrane action and the interpretation of D=11
supergravity as the effective field theory of a hypothetical supermembrane theory
\cite{bst,bps}. The subsequent demonstration \cite{dhis} that the GS action for
the IIA superstring is the `double-dimensional reduction' of the
D=11 supermembrane action suggested an interpretation of the IIA superstring as
a membrane wrapped around the circular 11th dimension. The case for this
interpretation was strengthened by the construction of the extreme membrane
solution of D=11 supergravity \cite{stelle}, together with the 
demonstration \cite{stelle,dgt} that it reduces in D=10 to the extreme string
solution of IIA supergravity, which had earlier been identified as the field
theory realization of the fundamental string \cite{dabhar}. A further important
development on this front was the construction of a fivebrane solution of D=11
supergravity \cite{guven}, which was subsequently shown to be geodesically
complete \cite{ght}. The fivebrane is the `magnetic' dual of the `electric'
membrane in D=11, in agreement with the general formula \cite{nep}that the dual of
a $p$-brane is a $\tilde p$-brane with $\tilde p= D-p-4$.
 
These connections between D=10 and D=11 physics were mostly classical; it still
seemed impossible that the {\it quantum} IIA superstring theory, with D=10 as
its critical dimension, could be 11-dimensional. In addition, the
non-renormalizability problem of D=11 supergravity appeared to have been 
replaced by the difficulty of a continuous spectrum for the first quantized
supermembrane \cite{dln}. An indication that the D=11 supermembrane might
after all be relevant to quantum superstring theory arose from consideration of
the soliton spectrum of compactified D=11 supergravity. It was noted that the
inclusion of wrapping modes of the membrane and fivebrane led to a spectrum of
solitons identical to that of the IIA superstring if, as could be argued on
other grounds, the latter includes the wrapping modes of the D=10 p-branes
carrying Ramond-Ramond charges \cite{hulltown}. But if this is taken to mean
that IIA superstring theory really is 11-dimensional then its non-perturbative
spectrum in D=10 must include the Kaluza-Klein excitations from D=11. These would
have long range ten-dimensional fields and so would have to appear as 
(BPS-saturated) `0-brane' solutions of IIA supergravity. Such solutions, and
their 6-brane duals, were already known to exist \cite{horstrom} and it was
therefore natural to interpret them as the field realization of the KK modes and
the KK 6-branes needed for the D=11 interpretation of IIA superstring 
theory \cite{pkta}. 

Because of the connection between the string coupling constant and the dilaton
it was clear that an improved understanding of the role of the dilaton would be
crucial to any advance in non-perturbative string theory. The fact that IIA
supergravity is the dimensional reduction of D=11 supergravity leads to a KK
interpretation of the dilaton as a measure of the radius $R_{11}$ of the 11th
dimension, and hence to a relation \cite{wita} between the string coupling
constant $g_s$ and $R_{11}$:
\begin{equation}
R_{11}= g_s^{2\over3}\ .
\label{eq:zeroa}
\end{equation}
This shows clearly that a power series in $g_s$ is an expansion
about $R_{11}=0$, so that the 11th dimension is indeed invisible in
string perturbation theory. In retrospect it is clear that that the connection
between $R_{11}$ and $g_s$ should have been exploited much earlier by
supermembrane enthusiasts. Ironically, the obstacle was the membrane itself; the
problem is that the area of a wrapped membrane, and hence its energy, is
proportional to $R_{11}$, leading one to expect the tension in D=10 to be
proportional to $R_{11}$ too. But if this were the case the tension of the
wrapped membrane would vanish in the $R_{11}\rightarrow 0$ limit. Thus it seemed
necessary to fix $R_{11}$ at some non-zero value, thereby precluding any
connection with {\it perturbative} string theory. What this overlooks is that the
energy as measured in D=10 superstring theory differs from that measured in D=11
by a power of $R_{11}$ which is precisely such as to ensure that the D=10 string
tension is {\it independent} of $R_{11}$, and hence non-zero in the
$R_{11}\rightarrow 0$ limit. This rescaling also ensures \cite{wita} that the
0-brane mass is proportional to $1/R_{11}$, as required for its
KK interpretation. 

In the strong coupling limit, in which $R_{11}\rightarrow\infty$, the vacuum is
11-dimensional Minkowski and the effective field theory is D=11 supergravity.
Some authors refer to this special point in the moduli space of vacua, or a
neighbourhood of it, as `M-theory', in which case superstring theories and
M-theory are on a somewhat similar footing as different approximations to the
underlying unified theory. This can be quite convenient but it leaves us without
a name for the `underlying unified theory'. One might continue to call it
`superstring theory' with the understanding that it is now non-perturbative, but
this terminology is inappropriate because strings do not play a privileged role
in the new theory even in vacua that are effectively 10-dimensional, and it is
membranes rather than strings that are important in the D=11 Minkowski vacuum.
Consequently, I shall adopt the other usage, in which M-theory {\it is} the
`underlying unified theory', i.e. M-theory is the quantum theory that unifies the
five superstring theories {\it and} D=11 supergravity. So `defined', M-theory is
still 11-dimensional in the sense that almost all its vacua  are 11-dimensional,
although some of these dimensions may be compact. The theory with the D=11
Minkowski vacuum will be called `uncompactified M-theory'. Superstring theories
can then be viewed as `compactifications of M-theory'. 

One might wonder how a {\it chiral} theory like the IIB superstring theory can 
be obtained by compactification of M-theory. This is a special case of a more
general problem of how chiral theories arise upon compactification from D=11,
given the no-go theorem for Kaluza-Klein (KK) compactification of D=11
supergravity \cite{Wit83}. It seems that there are two ways in which this no-go
theorem is circumvented by some M-theory compactifications, and both involve the
membrane or the fivebrane. One way stems from the fact that one can consider
compactifications of M-theory on orbifolds \cite{horwit,dasgup,witb,sen}, whereas
KK theory was traditionally restricted to manifolds. The other way, and the one
most directly relevant to the IIB superstring, is that chiral theories can
emerge as limits of non-chiral ones as a consequence of massive modes not
present in the KK spectrum. For example, D=11 supergravity compactified on $T^2$
consists of D=9 N=2 supergravity coupled to a KK tower of massive spin 2
multiplets. In the limit in which the area of the torus goes to zero, at
fixed shape, one obtains the (non-chiral) D=9 supergravity theory. In contrast,
M-theory compactified on $T^2$ also includes massive spin 2 multiplets coming from
membrane `wrapping' modes on $T^2$. These additional massive modes become
massless in the above limit, in such a way \cite{bho,asp,schwarz} that the
effective theory of the resulting massless fields is the {\it ten-dimensional}
and {\it chiral} IIB supergravity! Since this is a chiral theory there are two
equivalent versions of it, either left-handed or right-handed; which one we get
depends on the choice of sign of the Chern-Simons term in the D=11 supergravity
Lagrangian or, equivalently, the choice of sign of a related `Wess-Zumino' term in
the supermembrane action. Thus, M-theory incorporates an intrinsically `membrany'
mechanism that allows the emergence of chirality upon compactification. 

To examine this in more detail let us denote by $R_{10}$ and $R_{11}$
the radii of the torus in the M-theory compactification. The limit in which
$R_{10}\rightarrow \infty$, at fixed $R_{11}$, leads to the IIA theory with
coupling constant given by  (\ref{eq:zeroa}), i.e. 
\begin{equation}
g_s^{(A)}= R_{11}^{3/2}
\label{eq:zeroaa}
\end{equation}
where we now call the coupling constant $g_s^{(A)}$ to distinguish it from the
IIB coupling constant to be given below. For finite $R_{10}$ it is a result of
perturbative superstring theory \cite{dine,dbranes} that the IIA 
theory is equivalent to the the IIB theory compactified  on a circle of radius
$1/R_{10}$ (in units in which $\alpha'=1$); i.e. {\it the IIB theory is the
T-dual of the IIA theory}. It follows that the $S^1$-compactified IIB theory can
also be understood as $T^2$-compactified M-theory; the limit in which
$R_{11}\rightarrow 0$ and $R_{10}\rightarrow 0$, at a fixed ratio, then
leads to the uncompactified IIB theory with string  coupling constant
\begin{equation}
g_s^{(B)} = {R_{11}\over R_{10}}\ .
\label{eq:zerob}
\end{equation}
We may assume that $g_s^{(B)}\le 1$ since the interchange of $R_{10}$ and
$R_{11}$ is simply a reparametrisation of the torus. In other words, the IIB
theory at coupling $g_s^{(B)}$ is equivalent to the IIB theory at coupling
$1/g_s^{(B)}$. More generally, the discrete $Sl(2;Z)$ group of global
reparametrizations of the torus implies an $Sl(2;Z)$ symmetry of the IIB
theory, originally conjectured \cite{hulltown} on the basis of the $SL(2;R)$
symmetry of IIB supergravity \cite{twobe}. The way in which the IIA and the IIB
theories are found as $T^2$ compactifications of M-theory is shown in the
following diagram
\cite{asp}: 
\vskip 0.5cm
\epsfbox{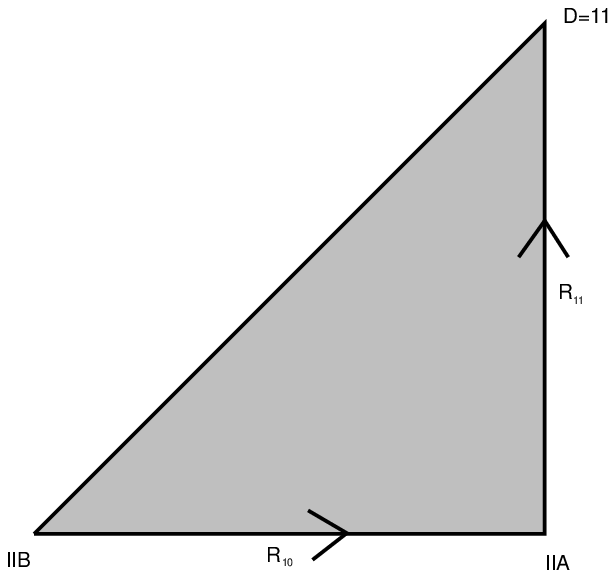} 
\vskip 0.5cm 
\noindent
A generic point on this diagram corresponds to an 11-dimensional vacuum. The
exceptions are those points with $R_{11}=0$ but non-vanishing $R_{10}$, 
corresponding to free string theories which may be ignored, and 
$(R_{10}, R_{11})= (0,0)$, corresponding to the uncompactified IIB
superstring theory, for which the vacuum is 10-dimensional Minkowski.
Actually, $(R_{10}, R_{11})= (0,0)$ is not really a `point' in the moduli space
because the IIB coupling constant depends on how the $(R_{10},
R_{11})\rightarrow (0,0)$ limit is taken. Thus while the
IIA theory has a more or less straightforward interpretation as an M-theory
compactification, as does the $S^1$-compactified IIB theory, the uncompactified
IIB theory is a singular limit of an M-theory compactification. 

Of course, it is equally true that the uncompactified M-theory is a limit of a
IIB `compactification' since, as the above figure illustrates, the
$S^1$-compactified IIB theory belongs to the same moduli space as
the $T^2$-compactified M-theory. In this sense, the two limiting theories are
on equal footing: neither is more fundamental than the other. 
In taking the IIB theory as the starting point for an exploration of the moduli
space of M-theory one might now wonder how the D=11 membrane could emerge from
the IIB theory. The answer lies in the fact that the IIB theory {\it is not just
a theory of strings}; as we shall see in some detail later it also contains other
objects, one of which is a 3-brane \cite{horstrom,dl,polb}. Upon $S^1$
compactification, the 3-brane can wrap around the circle to produce a membrane.
But this resolution of the puzzle raises another one; if the 3-brane does {\it
not} wrap around the circle then it remains a 3-brane and we now have to find
this object in the $T^2$-compactified M-theory. This is resolved by the fact that
the uncompactified M-theory is a theory not just of membranes but also, as
mentioned above, of fivebranes. A fivebrane wrapped around $T^2$ is a 3-brane. 
Of course, this fivebrane need not wrap around the $T^2$, so that we now have 
yet more branes to account for in the IIB theory. The end result of this type of
analysis is that all branes appearing in one compactification also appear in the
other \cite{schwarz}, as required by the equivalence of the two 
compactifications. Conversely, {\it it is precisely the existence of the various
branes that makes this equivalence possible}. 

It is clear from the description of IIB superstring theory as a limit of
$T^2$-compactified M-theory that the complex structure of the torus, viewed as a
Riemann surface, will survive the limit to become a parameter determining the
choice of IIB vacuum. In fact this parameter is the vacuum value of the
complex IIB supergravity field 
\begin{equation}
\tau = \ell + ie^{-\phi}\, ,
\label{eq:zeroc} 
\end{equation}
where $\phi$ is the scalar dilaton and $\ell$ is a pseuodoscalar `axion' field. 
If $\tau$ were assumed to be single-valued in the upper half plane then it 
would have to be constant over the compact KK space in any compactification of
the IIB theory. But, as its M-theory origin makes clear, $\tau$ actually takes
values in the fundamental domain of the modular group of the torus, so it need
not be single valued in the upper half plane. A class of compactifications that
exploits this possibility has been called `F-theory' \cite{vafa}. Since F-theory
has been associated with a hypothetical 12-dimensional theory, which would
appear to place it in an entirely different category, it is worthwhile to make a
detour to consider how F-theory fits into M-theory. An example will suffice.
D=11 supergravity, and hence (presumably) M-theory, can be compactified on 
a Ricci-flat four dimensional $K_3$ manifold \cite{dnptwo}. For some Ricci-flat
metrics, $K_3$ can be viewed as an elliptic fibration of $CP^1$, i.e. as a fibre
bundle where the fibre is a torus whose complex structure $\tau$ varies over
a Riemann sphere. Generically, there will be 24 singular points on the Riemann
sphere at which the torus degenerates but these are merely coordinate
singularities as long as no two singular points are coincident. Thus, there exist
M-theory compactifications on manifolds that are {\it locally} isomorphic to
$T^2\times S^2$. If the 2-torus is again shrunk to zero area we arrive at an $S^2$
compactification of the IIB theory in which the scalar field
$\tau$ varies over $S^2$. More generally, given a Ricci-flat manifold $E$ that
is an elliptic fibration of a compact manifold $B$, one can define
`F-theory on $E$' as IIB theory on $B$ with $\tau$ varying over $B$ in the way
prescribed by its identification as the complex structure of the torus in the
description of $E$ as an elliptic fibration. Formally, this would appear to
define `F-theory' as a 12-dimensional theory, but this is indeed purely formal.

Having seen how the Type II superstring theories are unified 
by M-theory, it remains for us to see how the superstring theories with only N=1
D=10 supersymmetry fit into this scheme. Firstly, we can ask how they are related
to each other. It is known \cite{narain,ginsparg} (at least to all orders in
perturbation theory) that the two heterotic string theories are related by
T-duality. The compactification of either theory on a circle allows a
non-vanishing `Wilson line' $\int\!\! A$ around the circle, where $A$ is the
Lie-algebra-valued gauge field of the effective supergravity/YM theory; this
amounts to choosing a non-zero expectation value for the component of $A$ in the
compact direction. This expectation value must lie in the Cartan subalgebra of
either $E_8\times E_8$ or $SO(32)$. Generically, this will break the gauge group
to $U(1)^{16}$ but special choices result in non-abelian groups, e.g.
$SO(16)\times SO(16)$. An
$SO(16)\times SO(16)$ heterotic theory obtained in this way by compactification
of the $SO(32)$ heterotic string theory on a circle of radius $R$ can be
similarly obtained by compactification of the $E_8\times E_8$ theory on a circle
of radius $1/R$. Thus, the uncompactified $SO(32)$ and $E_8\times E_8$ heterotic
string theories are theories with vacua that are limiting points in a single
connected space of vacua. We have already mentioned that the $SO(32)$ heterotic
and Type I theories are potentially equivalent non-perturbatively. We shall later
see some of the evidence for this. Anticipating this result, we see that there are
really only {\it two} distinct uncompactified D=10 superstring theories with N=1
supersymmetry, one with $SO(32)$ gauge group and one with $E_8\times E_8$ gauge
group. We shall call these the $SO(32)$ and $E_8\times E_8$ superstring theories.

Another clue from perturbative string theory is the fact that the Type I
theory is an `orientifold' of the Type IIB theory. The Type IIB string action is
invariant under a {\it worldsheet parity} operation,
$\Omega$, which exchanges the left and right movers. We can therefore find
a new string theory by gauging this symmetry \cite{horava,sagnotti}. This
projects out the worldsheet parity odd states of the Type IIB superstring
theory, leaving the states of the closed string sector of the Type I theory.
This sector is anomalous by itself, but one can now add an open string sector,
which can be viewed as an analogue of the twisted sector in the more
conventional orbifold construction. An anomaly free theory is found by the
inclusion of SO(32) Chan-Paton factors at the ends of open strings \cite{gsanom}.
This is the Type I string theory. By construction it is a theory of unoriented
closed and open strings. From its origin in the IIB theory it is clear that the
$S^1$-compactified Type I string must be 11-dimensional too. But since the IIA
theory has the more direct connection to D=11, and since this is the T-dual of
the IIB theory, we might expect to understand the 11-dimensional nature of the
Type I theory more readily by considering its T-dual, which is called the Type IA
(or Type I') theory \cite{dbranes}. 

The Type IA theory has some rather peculiar features. To understand them it is
convenient to start with the Type I theory in which the $SO(32)$ gauge group is
broken to $SO(16)\times SO(16)$ by the introduction of Wilson lines. Let
$Y(t,\sigma)$ be the map from the string worldsheet to the circle. T-duality
exchanges $Y$ for its worldsheet dual $\tilde Y$. Since $Y$ was a worldsheet
scalar, $\tilde Y$ is a pseudoscalar, i.e. 
\begin{equation}
\Omega [\tilde Y](t,\sigma)= -\tilde Y(t,-\sigma)\ .
\label{eq: omegy}
\end{equation}
Let $\tilde y$ be the constant in the mode expansion of $Y(\sigma)$; then $\Omega
[\tilde y] = -\tilde y$. The gauging of worldsheet  parity now implies that a
point on the circle with coordinate $\tilde y$ is identified with the point with
coordinate $-\tilde y$, so the circle becomes the orbifold $S^1/Z_2$ in which
the $Z_2$ action has two fixed points at $\tilde y=0,\pi$. In fact, since
$S^1/Z_2$ is just the closed interval ${\cal I}= [0,\pi]$, the fixed `points'
are actually 8-plane boundaries  of the 9-dimensional space, called
`orientifold' planes because the $Z_2$ action on $S^1$ is coupled with a
change of orientation on the worldsheet. 

Thus, the Type IA theory is effectively the Type IIA theory compactified on the
orbifold $S^1/Z_2$; closed strings that wind around the circle become open
strings stretched between two 8-plane boundaries, each of which is
associated with an $SO(16)$ gauge group. Actually, for reasons explained
below, the open strings in the Type IA theory do not end on the orientifold
8-plane boundaries {\it as such} but rather on 8-branes which happen to coincide
with them. Leaving this point aside for the moment, we are now in a position to
connect the N=1 superstring theories to M-theory. Since a IIA superstring is an
$S^1$-wrapped supermembrane in a D=11 KK spacetime, the open strings of the 1A
theory must be wrapped D=11 supermembranes stretched between two $S^1$-wrapped
9-plane boundaries of the D=11 KK spacetime. Let $L$ be the
distance between these boundaries and let $R$ be the radius of the circular
dimension, as measured in the D=11 metric. Then we can identify the Type
1A theory as the $R\rightarrow 0$ limit of M-theory compactified on a {\it
cylinder} of radius $R$ and length $L$. The stretched membrane described above is
effectively wrapped on the cylinder and has a closed string boundary on each of
the two ($S^1$-wrapped) 9-plane boundaries. Clearly, each string boundary must
carry an $SO(16)$ current algebra in order that an $SO(16)$ gauge theory emerge in
the $R\rightarrow 0$ limit. 

Suppose that we now increase $R$ at fixed $L$. The cylindrical D=11
supermembrane will eventually be transformed from a long tube of length $L$ and
small radius $R$ to a long strip of length $R$ and width $L$. For small $L$ the
two string boundaries of the supermembrane will appear as a single closed string
in a D=10 spacetime carrying an $SO(16)\times SO(16)$ current algebra. This is the
M-theory description of the heterotic string with $E_8\times E_8$ broken to
$SO(16)\times SO(16)$; taking $R\rightarrow \infty$ we recover the
uncompactifed D=10 heterotic string with unbroken $E_8\times E_8$ gauge group.
Its M-theory description is as a supermembrane stretched between two 9-plane
boundaries of the 10-dimensional space, separated by a distance $L$, with an
$E_8$ current algebra on each of its two string boundaries \cite{horwit}. The
string coupling constant turns out to be $g_s= L^{3\over2}$, so that the 11th
dimension, now taken to be the interval of length $L$, is indeed invisible in
perturbation theory. Of course, the limiting process just described
leads to an infinite heterotic string; a finite closed string has the
interpretation as a cylindrical D=11 supermembrane that is stretched between the
9-plane boundaries but is not otherwise wrapped. 

Let us now return to the $E_8\times E_8$ heterotic string theory compactified on a
circle of (large) radius $R$ with $E_8\times E_8$ broken to $SO(16)\times SO(16)$.
As we have seen, this has a description as M-theory compactified on a cylinder of
radius $R$ and length $L=g_s^{2/3}$. If $R$ is now continued from large to small
values, at fixed small $L$, we may switch to the T-dual description as an $SO(32)$
heterotic string compactified on a circle of radius $1/R$, with $SO(32)$
similarly broken to $SO(16)\times SO(16)$. The coupling constant of this theory
turns out to be $g^{het}_s= L/R$, which is still small as long as $R\gg L$. If we
continue to reduce $R$ we eventually move into the region for which
$R\ll L$. The heterotic string coupling constant is now large but we can
switch to the dual Type I description for which the string coupling constant is
\begin{equation}
g_s^I= 1/g_s^{het} = {R\over L} \ .
\label{eq:cchetone}
\end{equation}
In the limit in which both $R$ and $L$ go to zero at fixed small $g_s^I$ we
recover the uncompactified Type I theory from which we started. Thus, the moduli
space of M-theory compactified on a cylinder includes all superstring theories
with N=1 supersymmetry, as illustrated by the following figure
\cite{horwit}:
\vskip 0.5cm
\epsfbox{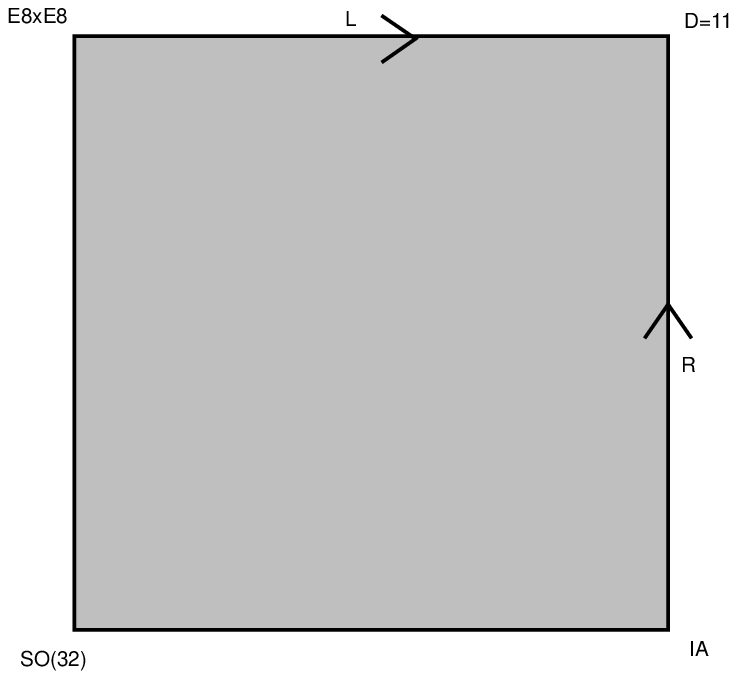} 
\vskip 0.5cm 
\noindent 
The generic vacuum in this moduli space is 11-dimensional but a 10-dimensional
theory with gauge group $SO(32)$ is obtained in the limit in which the cylinder
shrinks to zero area at fixed shape. 

We now return to the issue of open strings in the Type IA theory. T-duality
exchanges the Neumann boundary conditions on $Y$ at the ends of an open string to
Dirichlet boundary conditions on $\tilde Y$, i.e.
\begin{equation}
\partial_t {\tilde Y}(t,0) =0 \qquad \partial_t {\tilde Y}(t,\pi)=0\ .
\label{eq:zerof}
\end{equation}
It follows that open strings must now start at some fixed value of $\tilde Y$ 
and end  at some other, or the same, fixed value, i.e. open strings have their
ends tethered to some number of parallel 8-planes. Unlike Neumann boundary
conditions, Dirichlet boundary conditions do not prevent the flow of energy and
momentum off the ends of the string, so that the 8-planes on which the strings
end must be dynamical objects. They are called D-branes \cite{dbranes}, or
D-p-branes when we wish to specify the spatial dimension of the object. In this
case, the open strings end on D-8-branes. The N=2 supersymmetry of the IIB theory
is broken to N=1 in the Type I theory because of the restriction on the IIB
fermion fields at the ends of open strings. The N=2 supersymmetry of the IIA
theory is similarly broken to N=1 in the Type IA theory, but with the crucial
difference that since the ends of open strings lie in the D-8-branes {\it it is
only on these branes that the N=2 supersymmetry is broken; elsewhere, we have
have the unbroken N=2 supersymmetry of the IIA theory} \cite{polb}. Thus, the
Type IA theory is effectively equivalent to the IIA theory on an interval with
some number of D-branes. In fact, this number is 16, but to get an idea why we
shall need to understand some properties of D-branes \cite{polrev}.

It is typical of soliton solutions of supersymmetric field theories that
they carry conserved central charges of the supersymmetry algebra \cite{witol}.
The supersymmetry algebra then implies a bound on the mass, for fixed charge, that
is typically saturated by the soliton solution; the soliton is then said to be
`BPS-saturated'. BPS-saturated solitons preserve some fraction, often half,
of the supersymmetry of the vacuum. A further important, and related, feature 
is that the force between static BPS-saturated solitons vanishes so that there
also exist static, and BPS-saturated, multi-soliton solutions. There is a
generalization of all this to Type II supergravity theories in which a
`multi-soliton' is replaced by a solution representing a number of 
parallel infinite planar p-branes. The central charge in the supersymmetry
algebra becomes a p-form charge \cite{azcar}. Some of these p-branes, which all
preserve half the D=10 N=2 supersymmetry, carry the charges associated with
$(p+1)$-form gauge potentials coming from the Ramond-Ramond ($R\otimes R$) sector
of the Type II superstring theory; they are the $R\otimes R$ branes. There are
$R\otimes R$ p-branes of IIA supergravity for $p=0,2,4,6,8$ and $R\otimes R$
p-branes of IIB supergravity for $p=1,3,5,7$. As for p-branes in general, the long
wavelength dynamics is governed by an effective $(p+1)$-dimensional field theory,
but a feature peculiar to $R\otimes R$ branes is that this worldvolume field
theory includes a $U(1)$ gauge potential. This has a simple string theoretic
explanation: the  $R\otimes R$ branes of Type II supergravity theories are the
field theory realization of the Type II superstring D-branes, and the
(electric) $U(1)$ charges on the brane are the ends of open Type II strings. This
allows a string theory computation of the bosonic sector of the effective
worldvolume field theory \cite{leigh,li,douglas}; the full action is then
determined by supersymmetry and `kappa-symmetry' \cite{swedes,bergtown,agan}. The
result (upon partial gauge fixing) is a non-linear supersymmetric $U(1)$ gauge
theory of Born-Infeld type, except that the fields now depend only on the (p+1)
worldvolume coordinates of the brane. 

In the case of parallel multi D-branes there can be
open strings with one end on one brane and the other end on another brane.
Classically, such a string has a minimum energy proportional to the distance
between the branes. Supersymmetry ensures that this remains true
quantum-mechanically, so additional massless states can appear only when
two or more D-branes coincide. In fact, they {\it do} appear, and in just such a
way \cite{witbrane} that the $U(1)^n$ gauge group associated with $n$ coicident
D-branes  is enhanced to $U(n)$. Also, if this $n$ D-brane system
approaches an orientifold plane then further massless states (associated with
strings stretched between the D-branes and their mirror images) appear in just
such a way that $U(n)$ is enhanced to $SO(2n)$. The relevance of these
results to the Type IA theory is due to the fact noted above that this string
theory is just the Type IIA theory compactified on an interval of length $L$ with
some number of parallel D-8-branes. In the uncompactified theory we could allow
any number, $n$, of parallel D-8-branes. The number $n$ can be interpreted as the
total $R\otimes R$ charge; in general it will equal the number of branes minus
the number of anti-branes (although a configuration with both branes and
antibranes could not be static). On a compact `transverse' space, which is
one-dimensional in this instance, the total $R\otimes R$ charge must vanish, so
$n=0$ unless there are singular points of the compact space carrying non-zero
$R\otimes R$ charge. In compactification on $S^1/Z_2$ it turns out that the
orientifold planes each carry $R\otimes R$ charge $-8$, so that  precisely 16
D-8-branes are needed to achieve a vanishing total charge. To cancel the charge
{\it locally} we must put 8 branes on one orientifold plane and 8 on the other.
This leads to an $SO(16)$ gauge group associated with each orientifold plane;
this is the Type IA theory discussed above, i.e. the T-dual of the Type I theory
with $SO(32)$ broken to $SO(16)\times SO(16)$. 

It would be possible to arrange for the total 8-brane charge to vanish without it
vanishing locally by simply moving the 8-branes apart and/or away from the
orientifold planes. The generic configuration of this type would be one in which
the $SO(16)\times SO(16)$ symmetry is broken to $U(1)^{16}$; this is the T-dual
of the generic $S^1$ compactified Type I theory in which $SO(32)$ is broken to
its maximal abelian subgroup by an adjoint Higgs field. One could also move all
16 8-branes to one end in which case the gauge group would be enhanced to
$SO(32)$; this is the T-dual of the $S^1$ compactified Type I theory with
unbroken gauge group. In view of these possibilities, an obvious question is
why, in our earlier discusion, we needed to select the particular Type IA
theory with $SO(16)\times SO(16)$ gauge group. One might suppose that some other
configuration would be related to a version of M-theory in which 9-branes on the
D=10 boundaries of $S^1/Z_2$ compactified M-theory are moved away from the
boundary. However, a special feature of sources of 8-brane charge, i.e. 8-branes
or orientifold 8-planes, is that the dilaton grows with distance from the source
in such a way that the effective string coupling diverges at finite distance
\cite{witpol}. At fixed relative positions of the 8-branes the absolute distances
between them will grow with the distance $L$ between the orientifold planes so
that the 8-brane configuration is effectively constrained, as $L\rightarrow
\infty$, to approach the one in which the 8-brane charge is canceled locally.

We have now seen how all superstring theories are unified by a single theory,
M-theory, whose vacua are generically 11-dimensional. The moduli
spaces of vacua of the superstring theories with N=1 and N=2 supersymmetry are
connected by the special D=11 Lorentz invariant vacuum of uncompactified M-theory.
The connections between this D=11 uncompactified M-theory and the D=10
superstring theories are illustrated by the following figure (in which the two
N=1 string theories with $SO(32)$ gauge group are considered as a single
non-perturbative $SO(32)$ string theory):
\vskip 0.5cm
\epsfbox{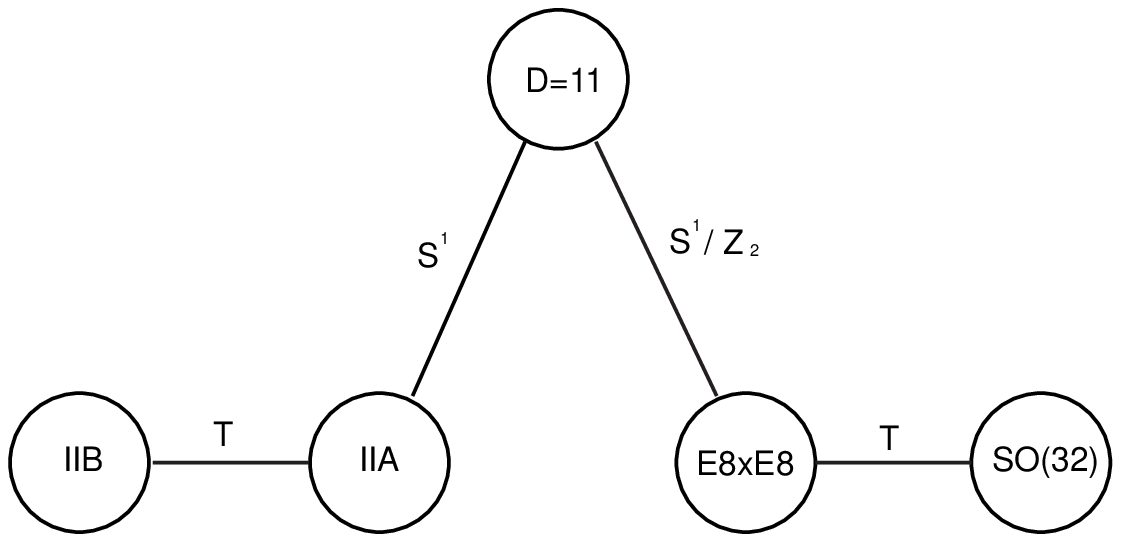} 
\vskip 0.5cm 
\noindent

This completes our overview of how superstring theories and D=11 supergravity 
are unified by M-theory, and why `branes' are crucial to this unification.
In the next lecture we backtrack to explain how considerations of D=10 and D=11
supersymmetry algebras both provide an explanation of why there are five
D=10 superstring theories and suggest a role for a D=11 supermembrane.
In the third lecture we shall see how the new scenario for superunification
via M-theory is supported by supergravity considerations. In the final lecture we
shall see how various features of D-branes, and other superstring p-branes, are
consequences of properties of the `M-branes' of M-theory. 

\section{Superstrings and the supermembrane}

It will be helpful to begin by reviewing how the various D=10
superstring theories arise. Our starting point will be the N=1 and N=2 
superspaces, which can be identified with the supertranslation groups. The
supertranslation algebras are spanned by the 10-momentum $P_\mu$ and one or
more Lorentz spinor charges. The minimal spinor in D=10 is both Majorana and
chiral. A Majorana spinor $Q$ is one for which $\bar Q= Q^T C$, where the bar
indicates the Dirac conjugate and $C$ is the antisymmetric real charge
conjugation matrix. There exists a representation of the Dirac algebra, the
Majorana representation, in which the Dirac matrices are real; in this
representation $C=\Gamma^0$ and a Majorana spinor is a real 32-component spinor.
A chiral spinor $Q_\pm$ is one for which
\begin{equation}
\Gamma_{11}Q_\pm = \pm Q_\pm
\label{eq:twoa}
\end{equation}
where $\Gamma_{11}$ is the product of all ten Dirac matrices. It satisfies
$(\Gamma_{11})^2=1$ and is clearly real in the Majorana representation, so 
chirality is compatible with reality in D=10. A chiral Majorana spinor has 16
independent real components. 

The N=1 supertranslation algebra is
\begin{equation}
\{Q^+_\alpha,Q^+_\beta\} = (C\Gamma^\mu {\cal P}^+ )_{\alpha\beta} P_\mu
\label{eq:twob}
\end{equation}
where $Q^+$ is a chiral Majorana spinor and ${\cal P}^+$ projects onto the
positive chirality subspace; the choice of positive or negative chirality is
of course purely convention.  An element of the supertranslation group is
obtained by exponentiation of the algebra element
\begin{equation}
X^\mu P_\mu + \bar\theta_+Q^+
\label{eq:twoc}
\end{equation}
where $X^\mu$ are the D=10 spacetime coordinates and $\theta_+$ is an anti-chiral
and anticommuting Majorana spinor coordinate. There are two N=2 supertranslation
algebras, according to whether the two supersymmetry charges have the same or
opposite chirality. If they have opposite chirality we can assemble them into a
single non-chiral Majorana charge $Q$. This leads to the IIA algebra
\begin{equation}
\{Q_\alpha,Q_\beta\} = (C\Gamma^\mu)_{\alpha\beta} P_\mu\ .
\label{eq:twod}
\end{equation}
If the two supersymmetry charges have the same chirality we can assemble them 
into the $SO(2)$ doublet $Q^{+\, I},\, (I=1,2)$ to arrive at the IIB algebra
\begin{equation}
\{Q^{+\, I}_\alpha,Q^{+\, J}_\beta\} = \delta^{IJ}
(C\Gamma^\mu{\cal P}^+)_{\alpha\beta} P_\mu\ .
\label{eq:twoe}
\end{equation}

With these supertranslation algebras in hand we can now turn to the
construction of superstring worldsheet actions in the Lorentz-covariant
Green-Schwarz (GS) formulation in which the fields are maps from the worldsheet
to superspace. We first introduce supertranslation invariant superspace 1-forms
on the three possible superspaces: 
\begin{equation}
\Pi^\mu = \cases{dX^\mu  - i\bar\theta_+\Gamma^\mu d\theta_+ & (heterotic) \cr
                 dX^\mu - i\bar\theta\Gamma^\mu d\theta & (IIA)\cr
dX^\mu - i\delta_{IJ}\bar\theta_+^I\Gamma^\mu d\theta_+^J & (IIB)\ .}
\label{eq:twof}
\end{equation} 
As indicated, the N=1 superspace case is relevant to the heterotic strings, the
Type I superstring being derived from the IIB superstring. Let
$\xi^i=(t,\sigma)$ be the worldsheet coordinates and let $\Pi_i^\mu$ denote
the 10-vector components of the induced worldsheet 1-forms. For example, in the
heterotic case we have
\begin{equation}
\Pi_i^\mu = \partial_i X^\mu -i\bar\theta_+\Gamma^\mu\partial_i \theta_+
\label{eq:twog}
\end{equation}
where $\{X^\mu(\xi),\theta^\alpha_+(\xi)\}$ are the worldsheet fields. Setting
the string tension to unity, for convenience, we can now write down the
supersymmetrized Nambu-Goto part of the superstring action,
\begin{equation}
S_{NG} = -\int d^2\xi\, \sqrt{-\det (\Pi_i\cdot\Pi_j)}\ .
\label{eq:twoh}
\end{equation}

For reasons reviewed elsewhere \cite{pktrev}, this is not the
complete action; it must be suplemented by a `Wess-Zumino term'. To construct it
we must search for {\it super-Poincar{\' e} invariant} closed forms on
superspace. This search reveals the following possibilities. Firstly, we have
some 3-forms 
\begin{equation}
h_{(3)}= \cases{\Pi^\mu\, d\bar\theta_+ \Gamma_\mu d\theta_+ & (heterotic)\cr
\Pi^\mu\, d\bar\theta \Gamma_\mu\Gamma_{11} d\theta & (IIA)\cr
\tilde S_{IJ}\, \Pi^\mu\, d\bar\theta_+^I \Gamma_\mu d\theta_+^J & (IIB)}
\label{eq:twoj}
\end{equation}
where $\Gamma_{11}$ is the product of the ten Dirac matrices $\Gamma^\mu$, and
$\tilde S_{IJ}$ are the entries of the $2\times 2$ matrix 
\begin{equation}
\tilde S =\pmatrix{1&0\cr 0&-1}\ .
\label{eq:twok}
\end{equation}
Secondly, we have the heterotic 7-form
\begin{equation}
h_{(7)}= \Pi^{\mu_1}\cdots \Pi^{\mu_7}
\, d\bar\theta_+\Gamma_{\mu_1\dots\mu_7}d\theta_+\ .
\label{eq:twol}
\end{equation}
The crucial fact about these forms is that they are closed (by virtue of Dirac
matrix identities valid in D=10) so locally we can write $h=db$. In fact, these
forms are exact, but this is a special feature of the chosen background. Given a
$(p+2)$-form $h_{(p+2)}$ we can find a super-Poincar{\' e} invariant Wess-Zumino
(WZ) type action for a p-dimensional object, i.e. a `p-brane', by integrating the
$(p+1)$-form $b_{(p+1)}$ over the $(p+1)$-dimensional worldvolume. Thus $h_{(7)}=
db_{(6)}$ could only be relevant to a 5-brane.  In fact, there is a heterotic
5-brane but our present concern is with the 3-forms (\ref{eq:twoj}) relevant to
1-branes, i.e. strings. Let $b_{ij}$ denote the components of the worldsheet 
2-form induced from $b$. We can now write down the Wess-Zumino part of the GS
superstring action:
\begin{equation}
S_{WZ} = {1\over2}\int d^2\xi\, \varepsilon^{ij}b_{ij}\ .
\label{eq:twom}
\end{equation}
The combined action 
\begin{equation}
S=S_{NG} + S_{WZ}
\label{eq:twon}
\end{equation}
has a fermionic gauge invariance, usually called `$\kappa$-symmetry', which allows
half the components of $\theta$ to be gauged away. On choosing a physical gauge
one finds that half of the original spacetime symmetries are {\it linearly
realized worldsheet supersymmetries}; without the $\kappa$-symmetry, they
would all be non-linearly realized \cite{hughes}. Thus, $\kappa$-symmetry is
essential for equivalence with the worldsheet supersymmetric NSR formulation of
superstring theory. This is one reason why the WZ term is essential to the
construction of a physically acceptable GS superstring theory, and it is this
that restricts the construction to N=1 and N=2 superspaces. While it is possible
to introduce supertranslation algebras for
$N>2$, a super-Poincar{\' e} invariant closed 3-form $h$ exists only for N=1 and
N=2. 

The Type II superstrings are closed strings whose covariant GS action is just
(\ref{eq:twon}), and for which the worldsheet fields are all periodic (recall
that the fermions are actually worldsheet scalars in this formulation; they
become worldsheet spinors only after gauge fixing the $\kappa$-symmetry). The
heterotic strings are closed strings based on the action (\ref{eq:twon}) for N=1
superspace, but conformal invariance of the first quantized string requires the
addition of a `heterotic' action
$S_{het}$ involving 32 worldsheet chiral fermions $\zeta^A,\, (A=1\dots 32)$. If
these are chosen to transform as half-densities then
\begin{equation}
S_{het} = {1\over2}\int d^2\xi\, \zeta_-^A \partial_+ \zeta_-^B\, \delta_{AB}
\label{eq:twoo}
\end{equation}
where $\partial_+$ is a chiral worldsheet derivative. Thus
\begin{equation}
S= S_{NG} + S_{WZ} + S_{het}
\label{eq:twop}
\end{equation}
is the GS action for the heterotic strings. The worldsheet fermions $\zeta^A$ 
may be periodic or anti-periodic so there are, a priori, many possible sectors
in the full Hilbert space of the first-quantized string. However, quantum
consistency requires a truncation to sectors with common boundary conditions
on groups of eight fermions, and further considerations along these lines leads
to the $SO(32)$ and $E_8\times E_8$ heterotic strings as the only ones
with a spacetime supersymmetric spectrum \cite{GSW}. This accounts for the Type
II and heterotic string theories. 

We have been considering these superstrings in a particular (Minkowski) 
background. More generally, any solution of the associated effective
supergravity theory provides a possible background, at least to leading order in
an expansion in powers of the inverse string tension $2\pi\alpha'$. We shall
discuss these supergravity theories in the following lecture. For the present it
will be sufficient to note that they all have in common the fields of N=1
supergravity, for which the bosonic fields are the metric
$g_{\mu\nu}$, an antisymmetric tensor gauge field $B_{\mu\nu}$ and a scalar
`dilaton' field $\phi$. We can regard the  background considered so far as one
for which $g$ is the Minkowski metric, $B$ vanishes and $\phi$ is constant.
Omitting worldsheet fermions, the worldsheet  action for a general background
involving these fields is 
\begin{equation}
S = -{1\over 4\pi\alpha'}\int\!\!d^2\xi\;\bigg\{
\sqrt{-\gamma}\gamma^{ij}g_{ij}  + \varepsilon^{ij} B_{ij}  + \alpha' \phi
\sqrt{-\gamma} R^{(2)}\bigg\}\ ,
\label{eq:threei}
\end{equation}
where we use here the `sigma model' formulation of the action in which
$\gamma_{ij}$ is an independent (but auxiliary) worldsheet metric with
scalar curvature $R^{(2)}$. The sigma model `coupling constants' 
$g_{ij}$, $B_{ij}$ and $\phi$ are the pullbacks to the worldsheet of the
spacetime fields. In the full action including worldsheet fermions the
2-form $B$ combines with the superspace 2-form $b$; both are part of the
complete WZ term.

For the heterotic string, the general bosonic background will also include a
background $SO(32)$ or $E_8\times E_8$ gauge potential. In the $SO(32)$ case
this background is easily accomodated by modifying $S_{het}$ in (\ref{eq:twop})
such that $\partial_+$ becomes a covariant derivative constructed from
the pullback of the spacetime gauge potential. In the $E_8\times
E_8$ case only an
$SO(16)\times SO(16)$ subgroup can be dealt with this way. For the Type II
strings there is a more serious omission, owing to the fact that spacetime bosons
in the string spectrum arise from two distinct sectors, the
Neveu-Schwarz/Neveu-Schwarz ($NS\otimes NS$) sector and the Ramond/Ramond
($R\otimes R$) sector. For present purposes we may define these sectors according
to whether the spacetime boson couples to a boson bilinear ($NS\otimes NS$) or to
a fermion bilinear ($R\otimes R$). The $R\otimes R$ fields are abelian
$(p+1)$-form potentials for various values of $p$ (as explained in the next
lecture) which couple to the worldsheet only through their $(p+2)$-form field
strengths (this is the only possibility compatible with gauge invariance). This
has the important consequence that $R\otimes R$ charges are {\it not} carried by
the Type II strings themselves. In contrast, since the 2-form potential $B$
couples `minimally' to the heterotic and Type II strings, they carry the charge
\begin{equation}
Q_1 = \int_{s^7}\! \star H\ ,
\label{eq:stringch}
\end{equation}
where (locally) $H=dB$ and $\star$ is the Hodge dual of spacetime. The
integral is over a 7-sphere surrounding the string, as shown schematically below:
\vskip 0.5cm
\epsfbox{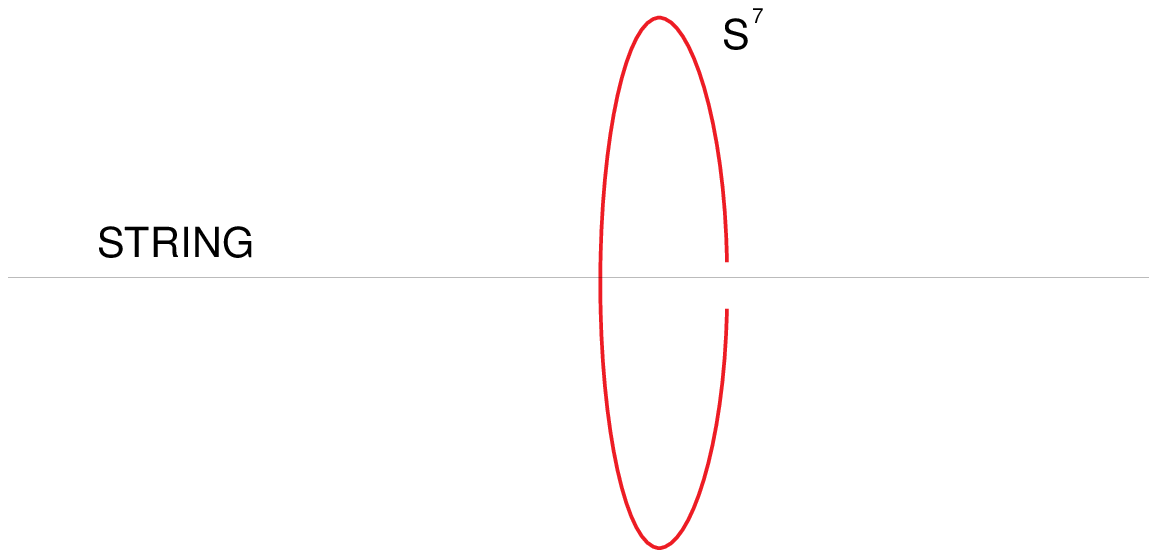} 
\vskip 0.5cm 
\noindent 
Because the heterotic and Type II strings are charged, in the sense that
$Q_1\ne0$, they cannot break; if this were to happen the 7-sphere could be slid
off the string and contracted to a point, which would imply $Q_1=0$. Actually,
this argument needs some qualification \cite{surgery} for Type II strings since
these can have endpoints on the D-branes which we encountered earlier.

We now turn to the remaining Type I superstring theory. Note that $\tilde S$ of
(\ref{eq:twok}) is {\it not} an $SO(2)$ invariant tensor, so the $SO(2)$
invariance of the IIB supertranslation algebra is broken by the IIB superstring
action. On the other hand, the minus sign in the definition of $\tilde S$ means
that the IIB action is invariant under a worldsheet parity operation $\Omega$,
induced by $\sigma\rightarrow -\sigma$, where the fields $X^\mu$ are assigned
positive parity (i.e. they are true scalars) and $(\theta_+^1\pm \theta_+^2)$ is
assigned parity $\pm 1$. That is, supressing Lorentz spinor and vector indices,
\begin{eqnarray}
\label{eq:twoq}
\Omega [ X](t,\sigma) &=& X(t,-\sigma)\cr
\Omega [(\theta_+^1 \pm \theta_+^2)](t,\sigma) &=& \pm 
(\theta_+^1 \pm \theta_+^2)(t,-\sigma) \ .
\end{eqnarray}
As mentioned earlier, the existence of this $Z_2$ symmetry means that we can 
find another superstring theory, the Type I theory, as an orientifold of
the IIB theory. The states of the closed string sector of the Type I theory are
found by projection onto the even worldsheet parity subspace of the Type IIB Fock
space. The $NS\otimes NS$ two-form $B$ is projected out since the worldsheet
boson bilinear to which it couples, $\varepsilon^{ij}\partial_i X^\mu\partial_j
X^\nu$, has odd parity. There are two parity-even fermion bilinears, both
involving the antisymmetric product of three Dirac matrices, but (as
explained in the next lecture) only one of them ($\bar\theta_+^1
\Gamma^{\mu\nu\rho}\theta_+^2$) couples to a 3-form field strength
from the $R\otimes R$ sector of the string theory. This is the only $R\otimes R$
field to survive the projection. Thus, the massless fields of the closed Type I
superstring are exactly those of N=1 supergravity, {\it but with the 2-form
gauge potential coming from the $R\otimes R$ sector}. 

An immediate consequence of this difference is that the Type I string does not
carry the charge $Q_1$ defined above and hence can break. In fact, the closed
string sector is anomalous by itself, but we can find an anomaly free theory by
the addition of an open string sector with $SO(32)$ Chan-Paton factors.
The states of the open string sector of the Type I theory are found by
quantization of the IIB worldsheet fields $Z^M(t,\sigma)$ subject to the 
constraint $Z=\Omega[Z]$. From (\ref{eq:twoq}) we see that this constraint
implies that
\begin{eqnarray}
\label{eq:twoqa}
X'(t,\sigma) &=& -X'(t,-\sigma)\cr
[\theta_+^1 - \theta_+^2](t,\sigma) &=& -
[\theta_+^1 - \theta_+^2](t,-\sigma) \ .
\end{eqnarray}
where the prime indicates differentiation with respect to $\sigma$.
This implies, in turn, that $X'=0$ and $\theta_+^1 = \theta_+^2$ at
$\sigma=0,\pi$, which are the standard boundary conditions at the ends of an 
open superstring.

It is now time to address the discrepancy between the symmetries of the IIB
superstring action and those of the IIB supertranslation algebra
(\ref{eq:twoe}). Recall that the latter has an SO(2) symmetry not shared by the
former. In fact, the discrepancy is illusory because the algebra
(\ref{eq:twoe}) is that relevant to the Minkowski vacuum. The algebra of
supersymmetry charges deduced as Noether charges of the IIB superstring action
contains an additional term arising from the fact that the WZ Lagrangian is not
invariant but changes by a total derivative. The algebra found this
way is \cite{azcar}
\begin{equation}
\{(Q^{+\, I}_\alpha,Q^{+\, J}_\beta\} =
\delta^{IJ}(C\Gamma^\mu {\cal P}^+)_{\alpha\beta} P_\mu +  \tilde S_{IJ}
(C\Gamma_\mu {\cal P}^+)_{\alpha\beta} Z^\mu
\label{eq:twobal}
\end{equation}
where $Z^{\mu}$ is the 1-form charge
\begin{equation}
Z^\mu =\oint \! dX^\mu\ ,
\label{eq:twobb}
\end{equation}
with the integral being taken over the image of the closed string in spacetime.
This charge is non-zero for strings that wind around a homology 1-cycle in space.
Effectively, this means that the charge $Z$ is relevant only for the
$S^1$-compactified IIB superstring, but one can then take the limit of infinite
radius to deduce that $Z$ is a 1-form charge carried by an infinite 
string in D=10 Minkowski spacetime. Thus, the supersymmetry algebra in the
presence of an infinite IIB superstring has the same symmetries as the
IIB superstring itself. 

Clearly a similar 1-form charge must appear in all the superstring theories for
which the worldsheet action contains a WZ term, i.e. all but the Type I
superstring. For example, for the IIA superstring we find that the algebra is
modified to
\begin{equation}
\{Q_\alpha,Q_\beta\} = (C\Gamma^\mu)_{\alpha\beta} P_\mu
+ (C\Gamma^\mu\Gamma_{11})_{\alpha\beta} Z_\mu\ .
\label{eq:twobc}
\end{equation}
It will prove instructive to rewrite the Type II algebras in terms of the
spinor charge $Q^+$ of the N=1 algebra and a second spinor charge $S^\pm$, where
$S^+$ is the second charge of the IIB algebra and $S^-$ of the IIA
algebra. In either case the supertranslation algebra is then 
\begin{eqnarray}
\label{eq:twobd}
\{Q^+_\alpha,Q^+_\beta\} &= (C\Gamma^\mu {\cal P}^+)_{\alpha\beta} (P+Z)_\mu \cr
\{S^\pm_\alpha,S^\pm_\beta\} &= (C\Gamma^\mu {\cal P}^\pm )_{\alpha\beta}
(P-Z)_\mu
\end{eqnarray}
where ${\cal P}^\pm$ are the projection operators onto the spinor subspaces of
positive or negative chirality. Note that the N=1 subalgebra (of $Q^+$) is
invariant under the interchange
\begin{equation}
P\leftrightarrow Z\ .
\label{eq:twobe}
\end{equation}
Of course, this symmetry is a classical one; in the quantum theory the spectrum
of $P$ and $Z$ as operators will generally be different, in which case
(\ref{eq:twobe}) would make no sense. For example the momentum in an
uncompactified direction can take any value while the corresponding winding
number has only one allowed value, zero. Suppose, however, that the $X^9$
direction is a circle of radius $R$. Then the spectrum of $P_9$ is isomorphic to
that of the winding number operator $Z^9$; the isomorphism involves the
exchange of $R$ with $1/R$ since the eigenvalues of $P_9$ are multiples of a unit
proportionl to $1/R$ while those of $Z^9$ are multiples of a unit proportional to
$R$. In fact, it is known that a heterotic string theory on a circle of radius
$R$ is equivalent to the same theory on a circle of radius $\alpha'/R$. This $Z_2$
symmetry of the heterotic string is called T-duality; it is actually a subgroup
of a much larger $SO(1,17;Z)$ discrete symmetry group of the generic
$S^1$-compactified heterotic string theory which is also called the T-duality
group \cite{giveon}. The invariance of the supersymmetry algebra under the 
interchange $P\leftrightarrow Z$ is clearly necessary for this to be possible.

If we had taken the N=1 supersymmetry algebra of the $S^\pm$ charges the
conclusion would have been the same, with the exchange symmetry being
$P\leftrightarrow -Z$, but the combined N=2 algebra has no analogous symmetry.
It follows that neither the IIA nor the IIB superstring, compactified on a
circle of radius $R$, is mapped to itself under the T-duality
transformation $R\rightarrow 1/R$. However, if we replace $S^\pm$
by $\Gamma^9 S^\mp$ at the same time that we make the exchange $P\leftrightarrow
Z$, then we recover the IIA algebra if we started with the IIB one, and
vice-versa (note that multiplication by $\Gamma^9$ maps a spinor of one D=10
chirality to one of the other chirality). In other words, the combined
transformation
\begin{equation}
P_9 \leftrightarrow Z^9 \qquad  S^\pm \leftrightarrow \Gamma^9 S^\mp
\label{eq:twobf}
\end{equation}
maps the IIA algebra into the IIB algebra, and vice versa. As
before this transformation makes sense in the quantum theory only if $X^9$ is the
coordinate of a circle, and then it must be accompanied by $R\rightarrow 1/R$.
In fact, it is known from perturbative string theory that the IIA and IIB
theories are interchanged by the T-duality transformation $R\rightarrow 1/R$. 

Finally, we turn to the D=11 supermembrane. A convenient starting point is again
the supertranslation algebra. In D=11 the minimal algebra is spanned by the
11-momentum $P_M$ and a 32-component Majorana spinor of the D=11 Lorentz group
$Q_\alpha$ obeying the anticommutation relation
\begin{equation}
\{Q_\alpha,Q_\beta\} = (C\Gamma^M)_{\alpha\beta} P_M\ .
\label{eq:mema}
\end{equation}
As before, we can introduce the supertranslation invariant 11-vector-valued
1-form on superspace
\begin{equation}
\Pi^M = dX^M -i\bar\theta\Gamma^M d\theta\ .
\label{eq:memb}
\end{equation}
We now search for super-Poincar{\' e} invariant closed forms on superspace. The
only possibility is the 4-form
\begin{equation}
h_{(4)}= \Pi^M\Pi^N\, d\bar\theta\Gamma_{MN}d\theta\ ,
\label{eq:memc}
\end{equation}
which leads us to expect a membrane rather than a string. The word `membrane'
has been used in the past to refer both to a generic p-brane and to a domain wall
in D spacetime dimensions, i.e. a $(D-2)$-brane. Here we use the word
`membrane' to mean exclusively a 2-brane 

The Nambu-Goto string action has an obvious p-brane generalization. The $p=2$,
i.e. membrane, case was first considered by Dirac, so this type of action is
sometimes called the Dirac action; its supersymmetric version (for unit surface
tension) is
\begin{equation}
S_D = -\int d^3\xi\, \sqrt{-\det (\Pi_i\cdot \Pi_j)}\ . 
\label{eq:memd}
\end{equation}
It turns out that there {\it is} a $\kappa$-invariant supermembrane action of
the form
\begin{equation}
S= S_D + S_{WZ}
\label{eq:meme}
\end{equation}
where $S_{WZ}$ is constructed from the 3-form $b_{(3)}$ for which $h_{(4)}=d
b_{(3)}$. As for the heterotic and Type II superstrings, the presence of the WZ
term implies a modification of the supersymmetry algebra. This time we find that
\begin{equation}
\{Q_\alpha,Q_\beta\} = (C\Gamma^M)_{\alpha\beta} P_M + 
(C\Gamma_{MN})_{\alpha\beta} Z_{(2)}^{MN} 
\label{eq:memf}
\end{equation}
where $Z_{(2)}$ is a 2-form charge. This D=11 supertranslation algebra can be
rewritten as a D=10 algebra by the simple expedient of splitting all charges into
their representations under the D=10 subgroup of the D=11 Lorentz group.
The D=11 supersymmetry charge becomes a D=10 Majorana spinor charge while
\begin{eqnarray}
\label{eq:memg}
P_M &=& (P_\mu, P_{11})\cr
Z_{(2)}^{MN} &=& (Z_{(2)}^{\mu\nu}, Z_{(2)}^{\mu\, 11} \equiv Z^\mu) \ .
\end{eqnarray}
In this new notation the algebra (\ref{eq:memf}) reads
\begin{eqnarray}
\label{eq:memh}
\{Q_\alpha,Q_\beta\} &= (C\Gamma^\mu)_{\alpha\beta} P_\mu
+ (C\Gamma^\mu\Gamma_{11})_{\alpha\beta} Z_\mu \cr
&+ (C\Gamma^{11})_{\alpha\beta}P_{11} +  (C\Gamma_{\mu\nu})_{\alpha\beta}
Z_{(2)}^{\mu\nu} \ .
\end{eqnarray}
Note the similarity to (\ref{eq:memf}), but in addition to the 1-form charge
$Z$ associated to the IIA string we also find a 0-form charge $P_{11}$ and a
2-form charge $Z_{(2)}$. This suggests not only that the IIA superstring is
really a D=11 supermembrane but also that the non-perturbative D=10 theory
is a theory not just of strings but also of 0-branes and 2-branes. This line of 
inquiry will be followed up in the last lecture.

We have seen in this lecture that the construction of string and membrane
actions with manifest spacetime supersymmetry requires the existence of a closed
3-form or 4-form on the relevant superspace, and that this requirement severely
restricts the possibilities. In fact, one can determine \cite{achu}, for any
spacetime dimension D and each N, the values of $p$ for which there exists a
closed superspace (p+2)-form (of the required dimension). The resulting table
of possibilities (the `old branescan') includes those D=10 and D=11 cases
discussed above, but it is now time to admit that the full story is rather more
complicated. It turns out that a {\it closed} $(p+2)$-form on superspace is
necessary only if (after gauge fixing the $\kappa$-symmetry) the worldvolume
fields consist exclusively of scalars and spinors. While this is the case for
p-brane solutions of flat space field theories, and for some p-brane solutions of
supergravity theories, it is not true in general, as was originally discovered by
an analysis of the small fluctuations about 5-brane solutions of Type II
supergravity theories \cite{chsa}. Examples in D=10 are provided by D-branes,
whose worldvolume field content is that of the D=10 vector multiplet
dimensionally reduced to (p+1) dimensions. Another example is the fivebrane
solution of D=11 supergravity, for which the field content is that of a
6-dimensional antisymmetric tensor multiplet. These  examples can be viewed as
having a common origin since the D=11 fivebrane is a type of M-theory D-brane in
the sense that it is an object on which a membrane can have a
boundary \cite{openpbrane,pktmbrane}. 

These additional possibilities for p-branes in D=10 and D=11 supergravity
theories are also associated with p-form extensions of the supersymmetry algebra.
For example, the D=11 superfivebrane is associated with a 5-form
extension of the D=11 superymmetry algebra. Thus, the full D=11
supertranslation algebra is \cite{pktdem}
\begin{equation}
\{Q_\alpha,Q_\beta\} = (C\Gamma^M)_{\alpha\beta} P_M + 
(C\Gamma_{MN})_{\alpha\beta} Z_{(2)}^{MN} + (C\Gamma_{MNPQR})_{\alpha\beta}
Z_{(5)}^{MNPQR}\, .
\label{eq:memftot}
\end{equation}
Note that the total number of algebraically independent charges that could appear
on the right hand side is 528. The number actually appearing is
\begin{equation}
11+55+462 =528
\label{count}
\end{equation}
so the algebra (\ref{eq:memftot}) is `maximally extended'. The three types of
charge apearing on the right hand side are those associated with the
supergraviton, the supermembrane and the superfivebrane, which are the three
basic ingredients of M-theory. It is therefore natural to regard
(\ref{eq:memftot}) as the `M-theory superalgebra'.

\section{Effective supergravities and strong coupling limits}

In this lecture we shall see how consideration of the D=10 and D=11
supersymmetry algebras, and the associated supergravity theories, essentially
determines the strong coupling limits of all uncompactified superstring theories.
Our starting point will be N=1 or N=2 D=10 superfields, which are superfunctions
of definite Grassman parity on the corresponding superspaces. The gauge-invariant
fields of D=10 supergravity theories are components of a single real scalar
superfield, subject to certain constraints. This description becomes quite
involved for the full non-linear theories but is simple at the
linearized level and a linearized analysis is sufficient to
reveal the field content. The restriction to N=1 and N=2 arises in this context
because the superfield expansion would otherwise contain high spin gauge fields  
with field equations that are consistent only in flat space.  

Consider first a real scalar superfield $\phi(X,\theta_+)$ on N=1 superspace.
This has the $\theta$-expansion \cite{Nilsson}
\begin{equation}
\phi(X,\theta_+) = \phi +  i\bar\theta_+\lambda^+ + 
i(\bar\theta_+ \Gamma^{\mu\nu\rho}\theta_+) H_{\mu\nu\rho} 
+ \dots
\label{eq:threea}
\end{equation}
The first component is a scalar, the `dilaton' $\phi$, followed by the dilatino
$\lambda^+$. Since $\theta_+$ has 16 components, the total number of
components at the $\theta^2$ level is 120, which is precisely the number of
components of the 3-form field $H$. The constraints on the superfield $\phi$ 
therefore occur at the $\theta^3$ level, where we find the gravitino field
strength. The Riemann tensor of the D=10 metric appears at the $\theta^4$ level,
and thereafter all higher dimension components are just derivatives of the lower
dimension ones. In addition, the constraints imply the Bianchi identity
$dH=0$, allowing us to write $H=dB$, where $B$ is a 2-form gauge potential. In
fact, the constraints also imply the field equations of $B$ and the other
fields in the graviton supermultiplet, of which the bosonic fields are
$(\phi, g_{\mu\nu}, B_{\mu\nu})$.

For N=1 we also have the possibility of a YM supermultiplet. The YM field
strength 2-form ${\cal F}$ is contained in a Lie-algebra valued anti-chiral
spinor superfield $\chi_+(X,\theta_+)$ with the
$\theta$-expansion
\begin{equation}
\chi_+(X,\theta_+) = \chi_+ + \Gamma^{\mu\nu}\theta_+ {\cal F}_{\mu\nu} +\dots 
\label{eq:threec}
\end{equation}
The constraints on this superfield, which occur at the $\theta^1$ level,
imply that there are no further independent components and that
${\cal F}$ satisfies both the YM Bianchi identity and field equation. When the 
YM multiplet is coupled to the graviton supermultiplet the superfield constraints
on both are modified in such a way that, {\sl inter alia}, the Bianchi
identity $dH=0$ is replaced by an `anomalous' one, equivalent to a modification of
$H$ to a include a YM Chern-Simons (CS) term \cite{cham}. Classically, N=1
supergravity can be coupled to a YM supermultiplet for any choice of the gauge
group $G$, but in the quantum theory cancellation of gravitational anomalies
requires $G$ to have dimension 496. If the group is non-abelian then there are
additional gauge and mixed anomalies that can be cancelled by the GS mechanism
only for $G=SO(32)$ or
$G=E_8\times E_8$, and then only by the inclusion of additional Lorentz CS terms.
Supersymmetry then requires the inclusion of an infinite number of further
higher-order interactions, and the full supersymmetric anomaly-free theory is not
known. This is a complicating feature of N=1 that is fortunately absent for N=2. 

We have now seen that the bosonic fields of the combined N=1 supergravity/YM
theory are 
\begin{equation}
(\phi, g_{\mu\nu}, b_{\mu\nu}; {\cal A}_\mu)\ ,
\label{eq:threee}
\end{equation}
where ${\cal A}$ is a YM 1-form taking values in the Lie algebra of $SO(32)$ or
$E_8\times E_8$. Omitting fermions (and neglecting higher-derivative terms in the
$\alpha'$ expansion) the action, for a particular choice of units, is 
\begin{equation}
S_{het} =  \int\! d^{10}x \sqrt{-g}\,
e^{-2\phi}\big[ R + 4|d\phi|^2 -{1\over3} |H|^2 - \alpha' {\rm tr} |{\cal
F}|^2\big]\ .	
\label{eq:threef}
\end{equation}
Notice that the constant vacuum value of $\phi$ is not fixed by the field
equations. In fact, the action is invariant under the dilation
\begin{eqnarray}
\label{eq:threeg}
\phi &\rightarrow \phi +4\lambda\cr
g_{\mu\nu} &\rightarrow e^{2\lambda} g_{\mu\nu}\cr 
b_{\mu\nu} &\rightarrow  e^{2\lambda} b_{\mu\nu}\cr
{\cal A}_\mu &\rightarrow e^\lambda {\cal A}_\mu\ ,
\end{eqnarray}
and this invariance extends to the full action. Since the vacuum, in which 
$\phi$ takes a particular value, is not invariant, this symmetry is a
spontaneously broken one for which the dilaton is the Nambu-Goldstone boson,
hence its name. 

We have chosen to write the action (\ref{eq:threef}) in a way that is
appropriate to the heterotic strings. Observe that the scalar curvature and all
other kinetic terms, appear multiplied by the factor $e^{-2\phi}$. It will be
important for what follows to understand why this factor is there. Observe that
for $\phi$ equal to its vacuum value  $\langle\phi\rangle$ the worldsheet action
(\ref{eq:threei}) includes the term \cite{FT} $-\langle\phi\rangle \chi$, where
\begin{equation}
\chi = {1\over4\pi} \int\! d^2\xi\, \sqrt{-\gamma}R^{(2)}
\label{eq:threej}
\end{equation}
is the worldsheet Euler number. For a closed Riemann surface of genus $g$ we
have
$\chi=2-2g$, so the Euclidean path-integrand $e^{-S}$ acquires a factor of
$g_s^{(2g-2)}$, where we have set
\begin{equation}
g_s = e^{\langle\phi\rangle}\ .
\label{eq:threek}
\end{equation}
Since the genus $g$ orders the perturbation series of closed string theories we
can identify $g_s$ as the closed string coupling constant. In particular,
classical closed string theory is associated with the Riemann sphere for which
$g=0$. This leads to a factor of $g_s^{-2}$ in the closed string effective
action, which is consistent with the $\phi$-dependence of the spacetime action
(\ref{eq:threef}).

Next, we turn to IIA supergravity. The gauge-invariant fields are again
contained in a single real scalar IIA superfield $\phi(X,\theta)$. Its
$\theta$-expansion is
\begin{eqnarray}
\label{eq:threel}
\phi(X,\theta) &=\phi +  i\bar\theta\lambda + i\bar\theta\theta M +
i\bar\theta \Gamma^{\mu\nu}\Gamma_{11}\theta K_{\mu\nu} + 
i\bar\theta \Gamma^{\mu\nu\rho}\Gamma_{11}\theta H_{\mu\nu\rho}\cr
& + \ i\bar\theta \Gamma^{\mu\nu\rho\sigma}\theta G_{\mu\nu\rho\sigma}  
+ \dots
\end{eqnarray}
Note that there are a total of 496 {\it possible} components at the $\theta^2$
level. In fact, only 376 appear, so there is a
constraint that sets to zero a 3-form field at the $\theta^2$ level in the
$\theta$-expansion. Apart from the gravitino field-strength and the Riemann
tensor there are again no further independent components. The constraints
also imply Bianchi identities for the field-strengths at the $\theta^2$ level. 
In particular, $dM=0$, so the scalar $M$ is just a constant. This is actually 
the cosmological constant of the `massive' IIA supergravity \cite{romans,brgpt}.
We shall set $M=0$ in these lectures. The other Bianchi identities imply that
$K=dC$ for 1-form potential $C$ (of KK origin in D=11), $H=dB$ for 2-form
potential $B$ and (at the linearized level) $G=dA$ for 3-form potential $A$.
Thus, the bosonic field content of IIA supergravity consists
of the fields of N=1 supergravity,
\begin{equation}
(\phi,g_{\mu\nu}, B_{\mu\nu})\ ,
\label{eq:threen}
\end{equation}
which are also those of the $NS\otimes NS$ sector of the IIA string theory,
together with the gauge potentials
\begin{equation}
(C_\mu, A_{\mu\nu\rho})\ ,
\label{eq:threeo}
\end{equation}
which are the fields from the $R\otimes R$ sector of the IIA string theory.
As in the N=1 case, the superfield constraints actually imply the full field
equations, but the bosonic field equations can also be derived from the
component action \cite{west}
\begin{eqnarray}
\label{eq:threep}
S_{IIA} &=  \int\! d^{10}x \bigg\{ \sqrt{-g}\,
e^{-2\phi}\big[ R + 4|d\phi|^2 -{1\over3} |H|^2\big] \cr
&- \sqrt{-g}\, \big[|K|^2 + {1\over12}|G|^2\big] \bigg\}\  + {1\over 144}\int
G\wedge G\wedge B \, ,
\end{eqnarray}
where $G= dA + 12B\wedge K$ is the non-linear version of the 4-form field
strength. 

Observe that the terms in (\ref{eq:threep}) involving the $R\otimes R$ fields
are {\it not} multiplied by a factor of $e^{-2\phi}$. Since this property 
of $R\otimes R$ fields has important consequences it deserves comment. In
the worldsheet supersymmetric NSR formulation of Type II string theories the
$R\otimes R$ fields do not couple to the string through local worldsheet
interactions but rather through bilinears of spin fields. These create cuts on
the Riemann surface which invalidate the conclusion we arrived at previously
that tree level closed string interactions are proportional to $g_s^{-2}$ (in
the GS formulation the RR fields {\it do} couple through local interactions
but the $\kappa$-symmetry makes the quantization of the GS
superstring problematic). Supersymmetry can be used to show that the $R\otimes
R$ fields must appear in the action as above, but it is also possible to show
this directly from string theory \cite{Bersiegel}. 

We turn now to IIB supergravity \cite{twobe}.  The field-strength superfield of
linearized IIB supergravity is again a real constrained scalar superfield
$\phi(X,\theta_+^I)$ with the $\theta$-expansion
\begin{eqnarray}
\label{eq:threeq}
\phi(X,\theta_+^I) &=\phi +  i\bar\theta_+^I\lambda_-^I +
\varepsilon^{IJ}\, i\bar\theta_+^I
\Gamma^\mu \theta_+^J\, L_\mu 
+i\bar\theta_+^I \Gamma^{\mu\nu\rho}\theta_+^J
\tilde H^{IJ}_{\mu\nu\rho}\cr
&+ \varepsilon^{IJ}\, i\bar\theta_+^I
\Gamma^{\mu\nu\rho\sigma\lambda}\theta_+^J
M^+_{\mu\nu\rho\sigma\lambda} + \dots 
\end{eqnarray}
The 5-form field $M^+$ is self-dual; this is not a constraint on the superfield 
but rather an automatic consequence of the chirality of the two $\theta$
coordinates. The tilde on $\tilde H$ indicates that this $SO(2)$ tensor is
tracefree, i.e. $\delta_{IJ}\tilde H^{IJ}=0$, or
\begin{equation}
\tilde H^{IJ} =\pmatrix{H&H'\cr H'&-H}\ .
\label{eq:hs}
\end{equation}
This {\it is} a constraint because, as in the IIA case, it means that of the
possible 496 components that could appear at the $\theta^2$ level only 376
actually do appear, {\it viz.} $L, H, H', M^+$. The superfield constraints imply
various Bianchi identities for these fields. In particular, $dL=0$, which
implies that $L=d\ell$ for pseudoscalar $\ell$, and (at the linearized
level) $dM^+=0$ which, because of the self-duality, implies not only that
$M^+=dC^+$ but also the linearized field equation for the 4-form $C^+$.
The other Bianchi identies imply that $H=dB$ and $H'=dB'$ for two 2-form
potentials $B$ and $B'$. Thus, the $NS\otimes NS$ fields of the IIB theory are
the same as those of the IIA superstring while the $R\otimes R$ fields are
\begin{equation}
(\ell, B'_{\mu\nu}, C^+_{\mu\nu\rho\sigma})\ ,
\label{eq:threer}
\end{equation}
where the superfix on $C^+$ is to remind us that its 5-form field strength
is self-dual. Note that we are regarding the pseudoscalar $\ell$ as a
gauge field here because it appears only through its field strength $L$. 

The self-duality of $M^+$ complicates the construction of an action for IIB
supergravity. There are some ways around this problem \cite{berkb,sorokin} but
they are rather unwieldy so we shall adopt the simpler procedure \cite{bbo} in
which the self-duality condition is temporarily dropped, thus allowing us to use
the standard Lagrangian for $C^+$. The self-duality condition is then simply
added to the field equations that follow from the variation of this action. With
this understanding, and omitting fermions, the IIB supergravity action is
\begin{eqnarray}
\label{eq:threes}
S_{IIB} &= \int d^{10}x {\sqrt {-g}}
\bigg\{e^{-2\phi}\big[ R + 4|d\phi|^2
-{1\over 3}|H|^2\big]  -2|d\ell|^2  \cr
&-{1\over 3}|H'-\ell H|^2  
 - {1\over 60}|M^+|^2\bigg\} -{1\over 48}\int C^+\wedge H\wedge H'\ ,
\end{eqnarray}
where the full non-linear Bianchi identity satisfied by $M^+$ is now
$dM^+ = H\wedge H'$. By combining this `modified' Bianchi identity with
the self-duality condition on $M^+$ we deduce that $d\star M^+ = H\wedge H'$,
which is just the $C^+$ field equation. Thus, the modification of the Bianchi
identity is needed for consistency  with
the self-duality condition. Notice that the $R\otimes R$ fields again appear in
the action without the factor of $e^{-2\phi}$.

We are now in a position to discuss the effective field theory of the Type I
superstring theory. The field content is necessarily the same as that of the
effective field theory of the $SO(32)$ heterotic string, but the 2-form
gauge potential is the field $B'$ from the $R\otimes R$ sector of the IIB theory,
so that its kinetic term must appear {\it without} the factor of $e^{-2\phi}$.
Moreover, since the YM fields couple to the string endpoints, their tree-level
amplitudes are associated with the disc (rather than the Riemann sphere) which
has Euler number equal to 1, and this leads to a factor of $e^{-\phi}$ (rather
than
$e^{-2\phi}$) multiplying the YM terms in the effective action. Thus, the bosonic
sector of the Type I effective action is (to leading order in an $\alpha'$
expansion)
\begin{equation}
S_I =  \int d^{10}x \sqrt{-g}\,\bigg\{
e^{-2\phi}\big[ R + 4|d\phi|^2\big] - e^{-\phi}{\rm tr} |{\cal F}|^2 - {1\over3}
|H'|^2\bigg\}\ .
\label{eq:threev}
\end{equation}

We have now found the bosonic sectors of the effective supergravity theories of
all five D=10 superstring theories. As we shall see, this provides a powerful
tool in the analysis of the possible strong coupling limits of these 
theories \cite{wita}. All that we need assume of a given superstring theory is
that it provides an asymptotic expansion in $g_s$ to some theory, or theories,
defined for all $g_s$. Given the existence of a non-perturbative theory, one can
continue $g_s$ from small to large values. Since $1/g_s$ is now small it is
reasonable to expect that the theory can now be approximated by {\it another}
asymptotic expansion in powers of $1/g_s$. What is this new perturbation theory?  
One can first ask what its massless sector will be. For sufficiently small $g_s$
the massless quanta are those of the initial superstring theory. One might imagine
that some of them could acquire masses as $g_s$ is increased, but
massless quanta can become massive only if their number, charges, and spins are
such that they can combine to form massive multiplets, which are all larger than
the irreducible massless ones. This condition is not met by the quanta
associated to the massless fields of any D=10 supergravity theory which must,
therefore, remain massless for all $g_s$, in particular for large $g_s$. Thus, the
only issue to  be addressed is whether any {\it other} massless quanta appear at
some non-zero value of $g_s$ (or as $g_s\rightarrow \infty$). 

Let us consider this question first for the Type IIB theory. {\it All} 
supermultiplets of massive one-particle states of the IIB supersymmetry algebra
contain states of at least spin 4. There are
some indications that higher-spin massless field theories might be
consistent if {\it all} spins are present but then only in the presence of
a cosmological constant. This makes it rather unlikely that additional massless
states could appear as the IIB string coupling constant is increased, so we
conclude that the massless states at strong coupling are almost certainly
the same as those at weak coupling. If so, the effective field theory at strong
coupling must again be IIB supergravity, since this is the only possiblity
permitted by supersymmetry. We can conclude that {\it there must exist a
symmetry of IIB supergravity which maps large negative $\phi$ to to large positive
$\phi$, i.e. small $g_s$ to large $g_s$}. 

In fact, there is such a symmetry \cite{twobe}. It is most
easily discussed in terms of the `Einstein-frame' metric 
\begin{equation}
g^{(E)}_{\mu\nu} =  e^{-{1\over2}\phi} g_{\mu\nu}\ ,
\label{eq:einmet}
\end{equation}
for which the action reads
\begin{eqnarray}
\label{eq:einmeta}
S^{(E)}_{IIB} &= \int d^{10}x {\sqrt {-g}}
\bigg\{ R  -2\big[|d\phi|^2 + e^{2\phi}|d\ell|^2\big] -{1\over 60}|M^+|^2 
-{1\over 3}e^{-\phi}|H|^2 \cr
&\qquad -{1\over3} e^\phi |H'-\ell H|^2 \big]\bigg\}
\ -{1\over 48}\int C^+\wedge H\wedge H'\ .
\end{eqnarray}
The action for $\phi$ and $\ell$ may now be recognised as that of a sigma-model
with target space $Sl(2;R)/U(1)$. The $Sl(2;R)$ group acts on $\phi$ and $\ell$ by
fractional linear transformations on the complex scalar 
$\tau = \ell + ie^{-\phi}$, i.e.
\begin{equation}
\tau \rightarrow {a\tau + b\over c\tau + d}\ ,
\label{eq:fraclin}
\end{equation}
where 
\begin{equation}
\pmatrix{a&b\cr c&d} \in Sl(2;R)\ .
\label{eq:fraclina}
\end{equation}
The full action (\ref{eq:einmeta}) is also $Sl(2;R)$ invariant provided that the
2-form-valued row vector $(B,-B')$ transforms as an $Sl(2;R)$ doublet:
\begin{equation}
\pmatrix{B'\cr B} \rightarrow \pmatrix{a&b\cr c&d} \pmatrix{B'\cr B}\ .
\label{eq:fraclinc}
\end{equation}
This $Sl(2;R)$ symmetry can be extended to the complete IIB supergravity action
including fermions. By choosing the special $Sl(2;R)$ matrix for which $a=d=0$ and
$b=-c=1$, we see that there is a symmetry of IIB supergravity
that takes $\phi\rightarrow -\phi$ for $\ell=0$ and is still such that large
negative $\phi$ is mapped to large positive $\phi$ for $\ell\ne0$, as predicted.

We have just seen that the symmetries of IIB supergravity are consistent with
the earlier deduction concerning the strong coupling limit of the IIB
superstring theory. We have not yet made any assumption about the
microscopic theory in this limit, but a now obvious guess is that the strongly
coupled IIB superstring theory is another IIB superstring theory \cite{wita}.
Ultimately, consistency of this guess implies that only a discrete $Sl(2;Z)$
subgroup of $Sl(2,R)$ can be realized as a symmetry of the non-perturbative IIB
superstring theory \cite{hulltown}; this discrete symmetry is itself
non-perturbative and therefore a surprise from the point of view of
conventional superstring theory. The embedding of $Sl(2;Z)$ in $Sl(2;R)$ depends
on the vacuum expectation value of $\ell$. When $\langle\ell\rangle=0$ the
$Sl(2;Z)$ group is the one for which the entries of the matrix
(\ref{eq:fraclina}) are integers; otherwise it is a similarity transformation of
an integer $Sl(2;R)$ matrix with the similarity transformation depending on
$\langle\ell\rangle$. Although the full $SL(2;Z)$ symmetry cannot be checked
directly, a $Z_2$ subgroup mapping weak coupling to strong coupling {\it must be
a symmetry of the full non-perturbative theory if this theory exists} because, as
we have seen,  this is required by supersymmetry. This $Z_2$ `duality' group is
precisely the one that takes $\phi\rightarrow -\phi$ when $\ell=0$. It is
instructive to note that this $Z_2$ transformation also takes $B$ to $B'$, and
hence maps the string charge $Q_1$ of (\ref{eq:stringch}) into a similar charge
$Q_1'$ defined with $B'$ replacing $B$. Thus the weak to strong coupling duality
of IIB superstring theory {\it requires the existence of a new type of string
carrying $R\otimes R$ charge}; this is just the D-string, to be discussed in the
following lecture. Given the existence of the D-string, T-duality implies the
existence of p-branes carrying {\it all} other $R\otimes R$ charges of either the
IIB or the IIA superstring theory, so the existence of the Type II D-branes
is a direct consequence of IIB superstring duality which is virtually a direct
consequence of the structure of the IIB supersymmetry algebra!

We turn now to the IIA theory. In the absence of additional massless fields
appearing for large $g_s$ the effective field theory at strong coupling would
have to be IIA supergravity again. But unlike IIB supergravity, there is no
symmetry that maps large positive $\phi$ to large negative $\phi$, so this
possibility is ruled out. It must be the case that additional massless fields
appear as $g_s\rightarrow \infty$. The main difference between IIA and IIB in 
the analysis of this question is that there is the possibility of a central
charge in the IIA algebra; as we saw from our earlier discussion of the IIA
superalgebra it has an interpretation as a KK charge. Centrally charged
multiplets can have maximum spin two and there {\it is} a (unique) consistent
coupling of IIA supergravity to massive centrally charged spin two
supermultiplets: it is the coupling determined by the compactification of D=11
supergravity to D=10. We conclude that the effective action at strong coupling
must be D=11 supergravity \cite{pkta,wita}. An immediate corollary is that, in
contrast to the IIB case, the strong coupling limit of the Type IIA superstring
theory {\it cannot be another superstring theory}.

The consistency of these conclusions can be checked by considering the
dimensional reduction of D=11 supergravity. Omitting fermions, the D=11
supergravity action is
\begin{eqnarray}
\label{eq:lecone}
S &=& {1\over \kappa^2}\int \! d^{11}x\,\bigg\{ \sqrt{-g}[R - {1\over
12}|F|^2] \cr
&& +\ {2\over (72)^2}\varepsilon^{M_1\dots M_{11}} F_{M_1\dots M_4} 
F_{M_5\dots M_8} A_{M_9 M_{10} M_{11}}\bigg\}\, ,
\end{eqnarray}
where $\kappa$ is the D=11 gravitational coupling constant. In general,
dimensional reduction to D=10 is possible once we assume that the D=11
background has a $U(1)$ isometry with Killing vector field $k$, such that the
4-form $F$ is also invariant, i.e ${\cal L}_k F=0$, where
${\cal L}_k$ is the Lie derivative with respect to $k$. In coordinates
$x^M=(x^\mu, y)$ for which $k=\partial/\partial y$, we can write the D=11
bosonic fields as
\begin{eqnarray}
\label{eq:lectwo}
ds^2 &=& e^{-{2\over3}\phi(x)}dx^\mu dx^\nu g_{\mu\nu}(x) +
e^{{4\over3}\phi(x)}\big( dy- dx^\mu C_\mu(x)\big)^2 \cr
A &=& {1\over 6} dx^\mu \wedge dx^\nu \wedge dx^\rho A_{\mu\nu\rho}(x) +
{1\over2}dx^\mu \wedge dx^\nu \wedge dy\, B_{\mu\nu}(x)\ ,
\end{eqnarray}
from which we can identify the D=10 bosonic fields. Note that they
coincide with the $NS\otimes NS$ fields $(\phi, g_{\mu\nu}, B_{\mu\nu})$ and the 
$R\otimes R$ fields $(C_\mu, A_{\mu\nu\rho})$ of IIA supergravity. Substituting
the Kaluza-Klein (KK) ansatz (\ref{eq:lectwo}) into the D=11 action
(\ref{eq:lecone}) leads precisely to the IIA action (\ref{eq:threep}) (in units
for which $R_{11}=\kappa^2$).

Since we have supposed that $k=\partial/\partial y$ is the Killing vector
field of a $U(1)$ isometry, the coordinate $y$ is periodically identified
and we may choose some standard identification without loss of generality, e.g.
$y\sim y+ 2\pi$. It then follows from ({\ref{eq:lectwo}) that the radius of the
11th dimension is $e^{{2\over3}\phi(x)}$. This is generally $x$-dependent but in
a KK vacuum we may set $\phi=\langle\phi\rangle$. In view of the relation
(\ref{eq:threek}) between the dilaton and the string coupling constant we
deduce that 
\begin{equation}
R_{11} = (g^{(A)}_s)^{2\over3}\ ,
\label{eq:radel}
\end{equation}
which is precisely the relation of (\ref{eq:zeroa}). This confirms
that the effective action of the IIA superstring theory in its strong coupling
limit is uncompactified D=11 supergravity, but it provides no clue to the nature
of the D=11 quantum theory for which this is the effective field theory.
One possibility is a supermembrane theory \cite{bst} because, as we shall explore
further in the next lecture, the IIA superstring transmutes at strong coupling
into a D=11 supermembrane. But one should distinguish between a superstring or
supermembrane {\it theory} and the superstring or supermembrane itself. The
absence of a dilaton in D=11 means that there is no small parameter in terms of
which one might define a perturbation theory, so it is not obvious that the
presence of a membrane in D=11 implies the existence of a supermembrane theory.
We shall return briefly to this point in the epilog to these lectures but it is
important to apreciate that, however things turn out, the major premise of
M-theory, for which the circumstantial evidence is now overwhelming, is that there
exists {\it some} consistent supersymmetric quantum theory in D=11 containing
membranes and fivebranes, with D=11 supergravity as its effective field
theory. 

As mentioned earlier, it is known that the IIA and IIB superstring theories are
equivalent, order by order in perturbation theory, after compactification on a
circle. Since this equivalence involves an interchange of KK modes and winding
modes it does not extend to the respective $S^1$-compactified supergravity
theories, but the massless modes in D=9 are unaffected by this
exchange so the D=9 N=2 supergravity obtained by dimensional reduction of IIA
supergravity must be equivalent to that obtained from IIB supergravity.
This also follows from supersymmetry because D=9 N=2 supergravity is unique up 
to field redefinitions, but to find the map from IIA fields to IIB
fields and vice-versa one must compare the two dimensionally reduced supergravity
theories. If we denote by $R_A$ and $R_B$ the radii of the circles in the $S^1$
compactified IIA and IIB supergravity theories, respectively, and by
$\phi_A$ and $\phi_B$ the respective dilatons, then one finds that
\begin{equation}
 e^{-\phi_A} R_A = e^{-\phi_B} \qquad R_A = 1/R_B\ .
\label{eq:tdual}
\end{equation}
Given that the IIA theory is $S^1$-compactified M-theory it follows that the
IIB theory can be found by a $T^2$ compactification, as discussed in the first
lecture. To determine the relation between the radius $R_{10}$ appearing in
that discussion and the radius $R_A$ of the IIA compactification we write
the 10-metric of IIA supergravity as $ds^2_{10} = ds^2_9 + R_A^2 (dx)^2$ where
$x$ is the coordinate of the circle (such that $x\sim x+2\pi$). Since $ds^2_{10}$
appears in the KK ansatz (\ref{eq:lecone}) with a factor of
$e^{-2\phi/3}$ we deduce that the radius of the circle from the D=11
perspective is 
\begin{equation}
R_{10} = e^{-{1\over3}\phi_A}R_A \, .
\label{eq:tduala}
\end{equation}
Combining this with (\ref{eq:tdual}) we have
\begin{equation}
e^{\phi_B} = e^{{2\over3}\phi_A}/R_{10}\, , 
\label{tdualb}
\end{equation}
and hence a formula for $g_s^{(B)}$ in terms of $g_s^{(A)}$ and $R_{10}$. Using 
(\ref{eq:radel}) to eliminate $g_s^{(A)}$ from this formula we find that
\begin{equation}
g_s^{(B)} = R_{11}/R_{10}\ ,
\label{eq:tdualc}
\end{equation}
which is precisely the formula (\ref{eq:zerob}). 

We now turn to the issue of the strong coupling dynamics of superstring theories
with N=1 supersymmetry. Let us consider first the Type I theory. At weak coupling
the effective field theory is (omitting fermions) given by (\ref{eq:threev}). We
must again address the question of whether additional massless fields can appear
at strong coupling. Because there is no possible central charge there is also no
possibility of a shortened massive supermultiplet, but because we now have only
N=1 supersymmetry the maximum spin of a massive supermultiplet could be as low as
two. As just explained, additional spin two fields becoming massless
signals the decompactification of an extra dimension. This is now
an unlikely possibility because the only higher dimensional supersymmetric
field theory is D=11 supergravity, which has double the required
number of supersymmetries and no gauge fields. Thus, the most likely
possibility is that the effective field theory at strong coupling is equivalent
to the one at weak coupling. This would mean that it must be obtainable by some
field redefinition that involves $\phi\rightarrow -\phi$. In contrast to the case
of IIB supergravity, there is no field redefinition of this type which takes the
effective field theory of the Type I superstring theory into itself, i.e. there
is no strong-to-weak coupling {\it symmetry}. However, there is a field
redefinition of this type that transforms the Type I effective field theory
into the $SO(32)$ heterotic effective field theory, (\ref{eq:threef}).
It is \cite{wita}
\begin{eqnarray}
\label{eq:threeu}
g_{\mu\nu} &&\rightarrow e^{-\phi}g_{\mu\nu}\cr
\phi &&\rightarrow -\phi\cr
B' &&\rightarrow B\cr
{\cal A} && \rightarrow \alpha' {\cal A}\ .
\end{eqnarray}
The equivalence of the two effective field
theories is not in itself surprising because the N=1 supergravity/YM theory is
unique up to field redefinitions once the gauge group is specified. However, the
fact that the required field redefinition involves a change of sign of the dilaton
is significant. It shows that {\it the strong coupling limit of the Type I string
theory is a theory with the same effective field theory as the $SO(32)$
heterotic string theory}. It is a now obvious guess that the strongly coupled
Type I string theory {\it is} the $SO(32)$ heterotic string theory, and
vice-versa. This is certainly the only possibility if the strong coupling limit of
one string theory is another string theory. Thus, subject to this assumption (for
which there is now plenty of additional evidence
\cite{hull,witpol}), the Type I and $SO(32)$ heterotic string theories are just
the weak and strong coupling expansions of a single non-perturbative `$SO(32)$
superstring theory'. 

It remains for us to determine the strong coupling limit of the D=10 
$E_8\times E_8$ heterotic string theory. In this case there is neither a 
weak-to-strong coupling symmetry of its effective field theory nor a
transformation that maps the latter into the effective field theory of any other
string theory. Thus, the strongly coupled $E_8\times E_8$ superstring theory
{\it cannot  be another string theory}. If the effective field theory at strong
coupling were  a KK theory it would have to be a compactification of D=11
supergravity, but the only conventional compactification is on $S^1$ and this
leads, as we have seen, to the IIA theory. The strong coupling limit of the
$E_8\times E_8$ superstring theory is therefore the most puzzling of the five. As
explained briefly in the earlier overview of M-theory unification, this puzzle is
resolved by the interpretation \cite{horwit} of the $E_8\times E_8$ superstring
theory as a compactification of M-theory on $S^1/Z_2$ .

\section{Branes from M-theory}

We have now seen how the picture sketched earlier in which all five
superstring theories are asymptotic expansions of a single 11-dimensional
theory, M-theory, is supported, and suggested, by the effective
supergravity theories. In fact, we have only just begun to mine the
information contained in these effective field theories. For example, much more
information is contained in the solutions admitted by them \cite{duffrev}. For
each (p+1)-form in the Lagrangian there is an associated electric-type p-brane
solution and a magnetic-type $(6-p)$-brane solution, carrying charges $Q_p$ and
$Q_{(6-p)}$ respectively. Actually, there are families of such solutions in which
the p-volume tension can be varied at will subject only to a BPS-type bound. For
reasons mentioned briefly in our M-theory overview, the solutions of most
interest are the `BPS-saturated' $p$-branes for which, as the name suggests, the
bound is saturated. The values of $p$ for which such solutions of a given theory
exist are given in the `M-theory brane-scan' of  Table 1. Only those p-branes with
$p\le6$ appear in electric/magnetic pairs, and  these will be the only ones to be
discussed here (the IIB 3-brane is an exception because it is self-dual, as
indicated by the `$+$' superscript). The IIB
7-brane and IIA 8-brane are included in the table only for the sake of
completeness: the IIB 7-brane is important for F-theory, the IIA 8-brane is
associated with the massive IIA theory. 
\begin{table}[t]
\caption{{\bf The M-Theory Branscan}}
\begin{center}
\begin{tabular}{||c||c|c|c|c|c|c|c||c|c||}
\hline\hline
{}{} &{} &{} &{} & {} & {} & {} & {} & {} & {} \\
D=11{} &{} &{}& 2 & {} & {} & 5 & {} & {} & {}\\
{}{} &{} &{} &{} & {} & {} & {} & {} & {} & {}\\
\hline
{}{} &{} &{} &{} & {} & {} & {} & {} & {} & {}\\
IIA {}& $0_D$ & $1_F$  &  $2_D$ & {} &  $4_D$ & $5_S$ & $6_D$ & {} & $8_D$\\ 
{}{} &{} &{} &{} & {} & {} & {} & {} & {} & {}\\
\hline
{}{} &{} &{} &{} & {} & {} & {} & {} & {} & {}\\
IIB&{}{} &$1_F$, $1_D$&{} {} & \ \ $3_D^+$\ \ & {}{} & $5_S$, $5_D$ & {}{} & \
$7_D$\  & {}{} \\ 
{}{} &{} &{} &{} & {} & {} & {} & {} & {} & {}\\
\hline
{}{} &{} &{} &{} & {} & {} & {} & {} & {} & {}\\
Type I {}&{} & $1_D$ & {} & {} & {} &  $5_D$  & {} & {} &{}\\ 
{}{} &{} &{} &{} & {} & {} & {} & {} & {} & {}\\
\hline
{}{} &{} &{} &{} & {} & {} & {} & {} & {} & {} \\
Het {}&{} & $1_F$ &{} &{} & {} & $5_S$ & {} & {} & {}\\ 
{}{} &{} &{} &{} & {} & {} & {} & {} & {} & {} \\
\hline\hline
\end{tabular}
\end{center}
\end{table}

The D=10 p-branes have been labelled in Table 1 with the subscript $F$, $D$, or
$S$, according to whether they are `Fundamental', `Dirichlet' or `Solitonic'. 
These adjectives are indicative of the string theory interpretation of the
various supergravity solutions. The Fundamental strings and Solitonic 5-branes
carry the electric or magnetic charges of the $NS\otimes NS$ 2-form potential
$B$, and are therefore present for all but the Type I theory. In particular,
we expect N=1 supergravity/YM solutions to represent the long range fields of
the heterotic string and its 5-brane dual, although their identification is not
straightforward in this case because of the Lorentz Chern-Simons terms required
for anomaly cancellation, and the consequent infinite series of higher derivative
terms then required by supersymmetry. Thus, for the heterotic string theories
one should rather seek massless field configurations that define conformal
field theories \cite{chsb}; these will be approximated by solutions of the
effective field theory. In contrast, the Type II supergravity p-brane solutions
define sigma-models with (4,4) worldsheet supersymmetry, which are automatically
conformally invariant. The Dirichlet branes are those carrying the $R\otimes R$
charges, which appear in all but the heterotic string theories. Their string
theory interpretation is in terms of open strings with mixed Dirichlet/Neumann
boundary conditions, as discussed in other contributions to the school
proceedings. Since the supergravity solutions will also be covered in
other contributions we shall not enter into details of them either.
For our purposes it will suffice to observe that the dependence of the p-volume
tension $T$ on the string coupling constant $g_s$, for the string-frame
metric, can be essentially read off from the supergravity Lagrangians 
given previously. The result is
\begin{equation}
T\sim \cases{1 & {\rm for a Fundamental string}\cr
1/ g_s & {\rm for a Dirichlet p-brane}\cr
1/g_s^2 & {\rm for a Solitonic 5-brane}\ . }
\label{branecoup}
\end{equation}
\noindent
Note that all but the `Fundamental' string are non-perturbative in $g_s$, as
required for consistency since there is no sign of any other extended object in 
string perturbation theory.

The chief purpose of these lectures is to show how M-theory unifies, and
encompasses, superstring theories. Although we made a start on this in
the previous lectures it should now be clear that part of our goal must be to
provide an M-theory explanation for all the superstring p-branes. We have
already seen some reasons for believing that the IIA superstring is a D=11
supermembrane wrapped around the 11th dimension, but Table 1 suggests that we
should also expect to be able to interpret the IIA D-4-brane as a wrapped D=11
fivebrane. Furthermore, properties of these IIA branes, e.g. their dependence on
the string coupling constant, should follow from properties of the D=11
branes, which we shall refer to collectively as `M-branes'. Our knowledge of
M-branes is rather limited at present but the effective worldvolume action for
the supermembrane is known and some features of the fivebrane action are also
known. 

Let us start with the supermembrane; the bosonic sector of its
worldvolume action is \cite{bst}
\begin{equation}
S= -{1\over 2\pi}\int\! d^3\xi \big\{ \sqrt{-\det g^{(11)}_{ij}}\  - \ {1\over6}
\varepsilon^{ijk}A^{(11)}_{ijk}\big\}
\label{eq:bosmem}
\end{equation}
where $g_{ij}^{(11)}$ and $A_{ijk}^{(11)}$ are pullbacks to the worldvolume of
the spacetime metric and 3-form of D=11 supergravity. The overall factor has
been chosen for later convenience. We shall take the spacetime fields to be of the
form given by the KK ansatz (\ref{eq:lectwo}), so that
\begin{eqnarray}
\label{eq:pulla}
g_{ij}^{(11)} &=& e^{-{2\over3}\phi}\partial_i X^\mu\partial_j X^\nu g_{\mu\nu}
+ e^{{4\over3}\phi}(\partial_i y - \partial_i X^\mu C_\mu)
(\partial_j y - \partial_j X^\mu C_\mu)\cr
A_{ijk}^{(11)} &=& \partial_i X^\mu\partial_j X^\nu \partial_k X^\rho
A_{\mu\nu\rho} + 3\partial_{[i}X^\mu\partial_j X^\nu \partial_{k]} y\; B_{\mu\nu}
\end{eqnarray} 
where the square brackets indicate total antisymmetrization
(with `strength one').
To obtain the action for a string in the D=10 background
provided by (\ref{eq:lectwo}) we must dimensionally-reduce the (2+1)-dimensional
supermembrane action to (1+1) dimensions. The standard dimensional reduction
ansatz would take all the worldvolume fields to be independent of one of the
worldvolume space coordinates, say $\rho$. This results in a $\rho$-independent
two-dimensional Lagrangian, but not one that can be identified with the usual
superstring Lagrangian. However, this ansatz is not appropriate for a membrane
wound around the 11th dimension. There is another way to achieve a
$\rho$-independent Lagrangian that makes use of the fact that a $U(1)$ isometry of
the D=11 background implies an invariance of the membrane action under the
transformation generated by the $U(1)$ Killing vector field $k$. Instead of
requiring the worldvolume fields to be
$\rho$-independent we can set 
\begin{equation}
\partial_\rho X^M = k^M(X) \ .
\label{eq:scherks}
\end{equation}
If we choose spacetime coordinates such that $k=\partial/\partial y$, where $y$
is the 11th coordinate, as in (\ref{eq:pulla}), then (\ref{eq:scherks}) reduces to
the condition that all worldvolume fields are $\rho$-independent except $y(\xi)$,
which is linear in $\rho$; we can then choose it to be proportional to $\rho$ by a
partial gauge choice. We can also choose the period of identification of
$\rho$ to be the same as that of $y$ without loss of generality. Thus,
(\ref{eq:scherks}) becomes
\begin{equation}
\partial_\rho X^\mu=0 \qquad y = \nu \rho 
\label{eq:dimreda}
\end{equation} 
for some integer $\nu$, which is the winding number of the
membrane around the $S^1$ factor of the D=11 spacetime. The choice $\nu=0$
corresponds to standard dimensional reduction while $\nu\ne 0$ corresponds to a
Scherk-Schwarz dimensional reduction, which is called `double-dimensional
reduction' in the context of worldvolume actions \cite{dhis}. 

Now let $\xi^i=(\sigma^\alpha,\rho)$. Using (\ref{eq:dimreda}) we then find that
the induced $3\times 3$ metric $g_{ij}^{(11)}$ is
\begin{equation}
g_{ij}^{(11)} = \pmatrix{ e^{-{2\over3}\phi}(g_{\alpha\beta} + 
e^{2\phi}C_\alpha C_\beta) & \nu e^{{4\over3}\phi}C_\alpha\cr
\nu e^{{4\over3}\phi}C_\beta & \nu^2 e^{{4\over3}\phi}}\ ,
\label{eq:dimredb}
\end{equation}
from which we compute
\begin{equation}
\sqrt{-\det g_{ij}^{(11)}} = \nu \sqrt{-\det g_{\alpha\beta}}\ .
\label{eq:dimredc}
\end{equation}
Similarly, (\ref{eq:dimreda}) implies that
\begin{equation}
{1\over 6} \varepsilon^{ijk} A^{(11)}_{ijk} = {1\over 2}\nu
\varepsilon^{\alpha\beta} B_{\alpha\beta}\ .
\label{eq:dimredd}
\end{equation}
The double-dimensionally reduced membrane action is therefore
\begin{equation}
S = -\nu\int d^2\sigma \big\{ \sqrt{-\det g_{\alpha\beta}}\  -
\ {1\over2}\varepsilon^{\alpha\beta}B_{\alpha\beta}\big\}\ ,
\label{eq:dimrede}
\end{equation}
but this is just $\nu$ times the string action (to leading order in $\alpha'$ and
with $2\pi\alpha'=1$) in the background provided by the NS-NS fields of IIA
supergravity. Applied to the full supermembrane action, the same procedure
yields the complete GS action for the IIA superstring in a general D=10 IIA
supergravity background.

Note that the string tension is proportional to $\nu$, which was to be expected
for a membrane wound $\nu$ times around a circle. Note also that, since all
$\phi$-dependence has cancelled from the action, the tension is
$g_s$-independent, as required for a `Fundamental' string. This is clearly a
special feature of the $3\times3$ matrix (\ref{eq:dimredb}). If we were to
double-dimensionally reduce the D=11 fivebrane action we would have a similar
calculation to perform but with a $6\times 6$ matrix. In this case we would find
that \begin{equation}
\sqrt{-\det g_{ij}^{(11)}} = \nu e^{-\phi}\sqrt{-\det g_{\alpha\beta}}\ .
\label{eq:fourbr}
\end{equation}
This is sufficient to show that the double-dimensionally reduced fourbrane
action will have a tension proportional to $1/g_s$, as required for its
interpretation as a IIA D-brane \cite{pkta}. 

We now have evidence that the IIA superstring and the IIA D-4-brane are just the
wrapped membrane and fivebrane of M-theory, but if we are to take this
seriously, we should consider all the implications.
For example, we cannot arbitrarily restrict the double dimensional reduction of
the supermembrane to a single choice of the winding number $\nu$. Clearly, the
$\nu=1$ string is the one that should be identified as the IIA superstring, and
it might seem that there is no place for the strings with higher winding numbers.
However, one should recall that a single charged field can create any number of
charged particles, which can appear as a single particle of higher charge if they
happen to be coincident. Similarly, a single string field can create any number of
coincident strings. It is a feature of supersymmetry that the force between these
strings is zero, so a superposition of $\nu$ unit tension strings would appear to
be a single string of tension $\nu$. In principle, we should also allow $\nu=0$.
The action (\ref{eq:dimrede}) vanishes when $\nu=0$ but a proper treatment
of the $\nu=0$ case leads to the action of a tensionless
string. This a potential difficulty because there is no place in IIA superstring
theory for a tensionless string. However, the $\nu=0$ string is {\it not} a
membrane wound around the compact 11th direction; it is rather a toroidal
membrane that has collapsed to a string. This is equally possible, in principle,
for a membrane in an uncompactified D=11 spacetime, so if there is indeed a
tensionless string it must already be present in D=11. But we are almost certainly
stepping outside the domain of validity of the supermembrane action when we
consider configurations of membranes collapsed to strings. 

The same caveat
applies to a membrane collapsed to a point, but let us nevertheless consider this
possibility. The action for such collapsed configurations, as deduced from the
supermembrane action itself, is that of the massless D=11 superparticle, for which
the bosonic part of the action, in a bosonic D=11 supergravity background can be
written in the (hamiltonian) form
\begin{equation}
S= \int dt \big[ \dot X^M P_M - {1\over2}\tilde v\; g^{MN} P_MP_N\big] 
\label{eq:sparticle}
\end{equation}
where $\tilde v$ is an independent worldline density and $P_M$ is the momentum
conjugate to $X^M$. In the supersymmetric case one finds that the states
of the quantum theory correspond to the massless fields of D=11 supergravity,
and this was one reason for thinking that D=11 supergravity might be the
effective field theory of a quantum supermembrane theory. As mentioned above,
the starting point of this approach is suspect because the supermembrane action
is likely to be only an effective one, but what we currently know about M-theory
requires us to postulate that its effective action is indeed D=11 supergravity.

Now, massive KK modes in D=10 can be interpreted as massless quanta of D=11 with
non-zero momentum in the compact direction, so the action (\ref{eq:sparticle})
can be used to determine some features of the KK spectrum of $S^1$ compactified
D=11 supergravity. The KK masses will be integral multiples of a unit proportional
to $1/R_{11}$, where $R_{11}$ is the radius of the circle, but this is the mass
unit as measured in the D=11 metric. To determine the mass in terms of the D=10
string-frame metric we choose the D=11 KK background of (\ref{eq:lectwo}) and set
$P_M= (P_\mu, P_y)$. Since $y$ is periodically identified with period $2\pi$,
the eigenvalues of its conjugate variable $P_y$ are integers. We
therefore set $P_y=n$ for integer $n$; the term $\dot y P_y$ is then a
total derivative which we may discard. Defining $v = e^{2\phi/3}\tilde v$, we
thereby arrive at the action
\begin{equation} 
S= \int dt\,\big\{ \dot X^\mu P_\mu  -
{1\over2}v \big[(P - nC)^2 - (ne^{-\phi})^2 \big]\big\}\ ,
\label{eq:spartca}
\end{equation}
which is that of a charged massive particle in a 10-dimensional spacetime.
Setting $\phi$ equal to its vacuum value $\langle\phi\rangle$ we see that the mass
$M$ and charge $Q$ of this particle are given by
\begin{equation} 
M= {n\over g_s} \qquad Q= n\ .
\label{eq:spartcb}
\end{equation}
As expected from its origin, the particle is charged with respect to the KK
vector field, which is the  $R\otimes R$ vector field of IIA superstring theory.
For $n=1$, its mass is precisely that required for identification as a
D-0-brane. Consideration of the complete D=11 massless
superparticle action \cite{bergtown} leads to an extension of (\ref{eq:spartca})
that includes a supersymmetry WZ term; this term implies an extension of the
supersymmetry algebra to include the charge $Q$ as a central charge. Standard
arguments can then be used to derive a BPS-type bound on the mass in terms of the
charge; this bound is saturated by (\ref{eq:spartcb}), which was to be expected
from the fact that KK modes are BPS-saturated. Actually, the D-0-branes provide
only the $n=1$ KK states whereas M-theory requires the existence of KK states for
each $n\ge1$. The $n\ge2$ states must appear as bound states in the $n$
D-0-brane system. The absence of forces between static D-0-branes implies
that these bound states must be at threshold. The issue of whether there exist
bound states at threshold is a delicate one and this prediction of M-theory
still awaits verification.

We are not yet finished with extracting the consequences of having replaced the
IIA superstring by a D=11 supermembrane. We have still to confront the most
obvious consequence of this idea. A membrane can as easily move in ten
dimensions as eleven so there must also exist a D=10 membrane in the
non-perturbative IIA superstring theory. We can determine its
effective action from that of the D=11 supermembrane
\cite{dufflu,pktmbrane,schm,bergtown}. Here we shall consider only the bosonic
action (\ref{eq:bosmem}). It will be convenient to rewrite this action in the
equivalent form
\begin{equation}
S= {1\over4\pi}\int d^3\xi\bigg\{ v^{-1}\det g^{(11)}_{ij} - v +
{1\over3}\varepsilon^{ijk}A^{(11)}_{ijk}\bigg\}
\label{eq:bosmema}
\end{equation}
where $v$ is an independent worldvolume density. The (classical) equivalence of
this action to (\ref{eq:bosmem}) follows by elimination of $v$ by means of its
Euler-Lagrange equation. As before we take the D=11 supergravity fields to be
given by the KK ansatz (\ref{eq:lectwo}). This implies that the induced fields
are those of (\ref{eq:pulla}), which we rewrite as
\begin{eqnarray}
\label{eq:tentoel}
g_{ij}^{(11)} &=& e^{-{2\over3}\phi} g_{ij} + e^{{4\over3}\phi}Y_i Y_j \cr
A_{ijk}^{(11)} &=& A_{ijk} + 3B_{[ij}Y_{k]} -3B_{[ij}C_{k]} \ ,
\end{eqnarray}
where $g_{ij}$, $A_{ijk}$ and $B_{ij}$ are the worldvolume fields induced by the
D=10 spacetime fields, and we have defined
\begin{equation}
Y \equiv dy + C\ .
\label{eq:ydef}
\end{equation}
It follows, since $g_{ij}$ is $3\times3$, that
\begin{equation}
\det g_{ij}^{(11)} = e^{-2\phi} \det [g_{ij} + e^{2\phi}Y_iY_j] \ .
\label{eq:detgij}
\end{equation}
Using properties of $3\times 3$ matrices we can rewrite this as
\begin{equation}
\det g_{ij}^{(11)}= (\det g_{ij})\big[ e^{-2\phi} + |Y|^2\big] \ ,
\label{eq:matid}
\end{equation}
where $|Y|^2 = Y_iY_j g^{ij}$. The action (\ref{eq:bosmema}) is then found to be
\begin{eqnarray}
\label{eq:bosmemc}
S &= {1\over4\pi}\int\! d^3\xi\, \big\{ v^{-1}e^{-2\phi}\det g_{ij} -v
 + {1\over3}\varepsilon^{ijk}[A_{ijk}- 3B_{ij} C_k ] \cr
& \qquad +\ v^{-1}(\det g_{ij}) |Y|^2 + \varepsilon^{ijk}B_{ij}Y_k\big\}\ . 
\end{eqnarray}

Note that the one-form $Y$ in the above action is just shorthand for the
expression in (\ref{eq:ydef}). As such, it satisfies the identity
\begin{equation}
d(Y -C)\equiv 0\ .
\label{eq:abosm}
\end{equation}
We can elevate $Y$ to the status of an independent field if we
impose this identity by a Lagrange multiplier. We can do this by adding to the
action (\ref{eq:bosmemc}) the term
\begin{equation}
-{1\over2\pi}\int\!  F\wedge (Y-C)
\label{eq:lagrangema}
\end{equation}
for closed two-form $F$. If $F$ were an independent field, it would be a Lagrange
multiplier for the constraint $(Y-C)=0$, whereas what we need is the weaker
constraint $d(Y-C)=0$. This constraint could be imposed by taking $F$ to be an
exact 2-form, i.e 
\begin{equation}
F=dV\ ,
\label{eq:borni}
\end{equation}
for some 1-form $V$ but this is slightly too strong a condition on $F$. If we
instead write $F=dV + 2\pi \omega^{(2)}$, where the closed 2-form $\omega^{(2)}$
belongs to an integral cohomology class of the membrane's worldvolume, then
(\ref{eq:lagrangema}) acquires the extra term 
\begin{equation}
\Delta S= \int\! \omega^{(2)} \wedge (Y-C)\ .
\label{eq:cohom}
\end{equation}
But the periodic identification of $y$ means that $dy/2\pi$, and hence 
$(Y-C)/2\pi$, also belongs to an integral cohomology class (of the worldvolume
after the pullback of forms from spacetime). Thus, $\Delta S/2\pi$ is an integer.
This implies that $\exp (i\Delta S)=1$ and hence that the addition to $F$ of
$2\pi\omega^{(2)}$ has no effect on the path-integral. We can take this freedom in
the definition of $F$ into account by allowing the 1-form gauge potential $V$ to
be defined only locally, such that the flux of $F/2\pi$ over any 2-cycle is an
integer. This is equivalent to the statement that $iV/2\pi$ is a $U(1)$ gauge
potential (as against merely an abelian one).   

Now that we have settled the question of the nature of the gauge potential $V$
introduced by the Lagrange multiplier term (\ref{eq:lagrangema}) we add this term
to (\ref{eq:bosmemc}) to obtain the equivalent action 
\begin{eqnarray}
\label{eq:bosmemd}
S &= {1\over8\pi^2}\int\! d^3\xi\, \bigg\{ v^{-1}e^{-2\phi}\det g_{ij} -v
 + {1\over3}\varepsilon^{ijk}\big[A_{ijk}+ 3{\cal F}_{ij}C_k\big] \cr
& +\ v^{-1}(\det g_{ij}) |Y^2| - 
\varepsilon^{ijk}{\cal F}_{ij} Y_k \bigg\}\ , 
\end{eqnarray}
where we have defined the `modified' field strength
\begin{equation}
{\cal F}_{ij} = F_{ij} - B_{ij}\ .
\label{eq:moddeff}
\end{equation}
Here, in accordance with our condensed notation, $B$ should be understood to be
the pullback to the worldvolume of the spacetime 2-form potential $B$.
Since $Y$ is an independent field in the new action, it can be eliminated by its
algebraic, and linear, Euler-Lagrange equation
\begin{equation}
Y^i= {v\over2\det g} \varepsilon^{ijk} {\cal F}_{jk}\ .
\label{eq:yeq}
\end{equation}
The resulting action can then be simplified by use of the $3\times 3$
matrix identity 
\begin{equation}
\det [ g_{ij} \pm J_{ij}]\equiv (\det g_{ij}) \big[ 1+ {1\over2}|J|^2\big]
\label{eq:ideng}
\end{equation}
where $J_{ij}$ is any antisymmetric matrix and $|J|^2= g^{ij}g^{kl}J_{ik}J_{jl}$.
The result of these manipulations is
\begin{eqnarray}
\label{eq:bosmeme}
S &=  {1\over8\pi^2}\int d^3\xi\, \bigg\{ -\tilde v\, e^{-2\phi}
+\tilde v{}^{-1}\det(g_{ij} +{\cal F}_{ij}) \cr
&\qquad +\ {1\over3}\varepsilon^{ijk}\big[A_{ijk}  + 3 {\cal
F}_{ij}C_k\big]\bigg\}\ .
\end{eqnarray}
where $\tilde v=-\det(g_{ij})/v$. Finally, elimination of $\tilde v$ yields
\begin{equation}
S =  -{1\over4\pi^2}\int d^3\xi\, e^{-\phi}\sqrt{- \det (g_{ij}+ {\cal F}_{ij})}\ 
+\  {1\over4\pi^2}\int_w\! (A + {\cal F}\wedge C)\ .
\label{eq:bosmemf}
\end{equation}
where the final `Wess-Zumino' term has now been written as an integral of a
3-form over the worldvolume $w$. The dependence on ${\cal F}$ in the first term is
reminiscent of the Born-Infeld action for `non-linear electrodynamics', so the
full action is called the Dirac-Born-Infeld (DBI) action. Setting $\phi$ to its
vacuum value we see that the 2-brane tension is proportional to $1/g_s$, as
required for its interpretation as the IIA D-2-brane.  

We have now shown that the D=11 supermembrane action requires the D-2-brane of
IIA superstring theory to have an effective worldvolume action of the form
\begin{equation}
S = S_{DBI} + S_{WZ}
\label{eq:bosmemh}
\end{equation}
where $S_{DBI}$ is the DBI action with tension of order $1/g_s$, and $S_{WZ}$ is
a `Wess-Zumino' term. This `prediction' of M-theory is verifiable by a string
theory calculation, which also shows that it is a general feature. The WZ term
provides the coupling of the D-brane to the $R\otimes R$ fields. The
`leading' term in the WZ term is always of the form $\int\! C^{(p+1)}$, i.e. a
minimal coupling of the D-brane to the $R\otimes R$ potential $C^{(p+1)}$,
implying that the D-brane is a charged source for $C^{(p+1)}$. We have now seen
two examples of this: the D-0-brane, which is a source for the 1-form potential
$C^{(1)}=C$, and the D-2-brane, which is a source for the 3-form potential
$C^{(3)}=A$. In the latter case, a magnetic source of $F$ on the D-2-brane is
also a source of $C$; this has some interesting implications but we
shall have to pass over them here.

Just as the existence of a membrane in D=11 implies the existence of one in
D=10, the existence of a fivebrane in D=11 implies the existence of a D=10
5-brane. Returning to (\ref{eq:tentoel}) but interpreting the induced metric as
one on the 6-dimensional worldvolume of a fivebrane we have
\begin{equation}
\det g_{ij}^{(11)} = e^{-4\phi} \det [g_{ij} + e^{2\phi}Y_iY_j] 
\label{eq:detgijk}
\end{equation}
in place of (\ref{eq:detgij}). Determination of the full D=10 5-brane action
from M-theory is complicated by the fact that the D=11 fivebrane action is not
yet fully known; its worldvolume fields include a 2-form potential with
self-dual 3-form field strength \cite{gibtown}. Nevertheless, its Lagrangian will
include a term of the standard Dirac form and this, together with
(\ref{eq:detgijk}), is sufficient to show that the tension of the D=10 5-brane is
$1/g_s^2$, as expected from its `Solitonic' interpretation in string theory. We
have still to consider the D-6-brane and the D-8-brane. The D-6-brane does not
have an M-brane interpretation, although it does have a simple M-theory
interpretation \cite{pkta} as a generalized KK monopole. The M-theoretic
interpretation of the D-8-brane is currently problematic since its long range
fields solve the equations of the `massive' IIA supergravity which, as far as we
can see, cannot be obtained from D=11 supergravity. Hopefully, this mystery will
be cleared up in the near future. In any case, the M-theory predictions agree with
results obtainable from IIA superstring theory in so far as it is currently
possible to check.   

We turn now to the IIB branes. Their worldvolume actions can be determined
indirectly from M-theory by virtue of the fact that the IIB theory and the IIA
theory are T-dual. For example, if the D-2-brane action given above is
compactified on $S^1$ and the IIA background is replaced by its T-dual IIB
background then we obtain the action for the D-1-brane, or D-string. If this
D-string action is compactified on $S^1$ and the IIB background is replaced by
the original IIA background then we recover the D-0-brane action. This last step
provides the simplest illustration of the procedure, so we shall consider some of
the details \cite{bachas,bergroo}. To do this we must depart slightly
from the logic in which the D-brane actions are derived from M-theory by first
postulating the (bosonic sector of the) D-string action and then showing that it
leads to the same D-0-brane action as we previously derived from M-theory.
Actually, in order to illustrate an additional point we shall
start with the action for a D-string with an integer $n$ times the tension of a
single D-string. This is
\begin{equation}
S = -{n\over 2\pi} \int\! d^2\sigma\,\big\{ e^{-\phi_B}
\sqrt{-\det(g_{ij} + {\cal F}_{ij})}\ + {1\over2}\varepsilon^{ij}(B'_{ij} +
\ell {\cal F}_{ij}) \big\}  
\label{eq:dstring}
\end{equation}
where ${\cal F}$ is the `modified' 2-form field strength introduced in 
(\ref{eq:moddeff}). We shall proceed by first converting this
action to Hamiltonian form. The procedure for doing this is standard so we go
straight to the final result, which is
\begin{eqnarray}
\label{eq:hamstring}
S &=& -{1\over 2\pi} \int\! dt \int_0^{2\pi}\!\! d\sigma\, \Bigg\{ \dot X\cdot P +
\dot V_\sigma E + V_t E' + s\; X'\cdot P   \cr  
&&\! -{1\over2}v \bigg[(P - n{\cal B}' + E{\cal B})^2
+ (X')^2[ (E-n\ell)^2 +n^2e^{-2\phi_B}] \bigg] \Bigg\}
\end{eqnarray}
where
\begin{eqnarray}
\label{eq:loopa} 
{\cal B}_\mu &=& (X')^\nu B_{\mu\nu}\cr
{\cal B}'_\mu &=& (X')^\nu B'_{\mu\nu}\ ,
\end{eqnarray}
and the Lagrange multipliers $v$ and $s$ are the analogues of the lapse and
shift functions of General Relativity. The variables $P_\mu$ and $E$ are the
conjugate momenta to $X^\mu$ and $V_\sigma$, respectively. Thus $E$ is effectively
the BI electric field and the constraint imposed by $V_t$ is the 1+1 dimensional
version of the usual Gauss' law constraint of electrodynamics. Note that a
prime is used to denote differentiation with respect to $\sigma$ except in
$B'$ where it distinguishes the $R\otimes R$ 2-form potential from the $NS\otimes
NS$ one.  

With a view to double dimensional reduction we now suppose that the IIB background
is of KK type, i.e. admits a $U(1)$ Killing vector field $k=\partial/\partial u$.
If $u$ is identified with period $2\pi$ then the radius of the
compact direction is $R_B=\sqrt{k^2}$. We then take all worldsheet fields to be
$\sigma$-independent with the exception of $u$, which we set equal to $\sigma$.
The constraint imposed by the `shift' function $s$ now reduces to
$k\cdot P=0$, so the use of this constraint removes the conjugate pair
$(u,k\cdot P)$ from the action. Moreover, since the Lagrangian is now
$\sigma$-independent, the $\sigma$ integration can be trivially done, leading to
a factor of $2\pi$. At this point, we have a particle action in a background
provided by the fields of IIB supergravity, but we may now use the `T-duality
rules' to express the background in terms of IIA fields. We have already come
across a subset of these rules in (\ref{eq:tdual}). These can be extended to the
full set of IIA and IIB supergravity fields \cite{bho}. We shall not go into the
details here except to say that $({\cal B}-\ell{\cal B}')$  becomes the IIA KK
1-form $C$; the net result of using the T-duality rules in the
double-dimensionally reduced D-string action (\ref{eq:hamstring}) is 
\begin{equation}
S = -\int\! dt \Big\{ \dot {\tilde X}\cdot \tilde P -{1\over2}v 
\big[(\tilde P - nC) ^2 + n^2e^{-2\phi_A}\big] \Big\}
\label{eq:loopc}
\end{equation}
where $\tilde X^\mu = ( X^{\bar\mu}, V_\sigma)$ with $\bar\mu=0,1,\dots,8$
and $\tilde P_\mu = (P_{\bar\mu},E)$. The IIA metric is also of KK form with
Killing vector field $\tilde k=\partial/\partial V_\sigma$, and $\tilde k^2=
R_A^2$. This is precisely the D-0-brane action (\ref{eq:spartca}) in a D=10 
KK background provided that $R_A$ can be identified as the radius of the
compact 10th dimension, which it can be if $V_\sigma$ is an angular variable with
period $2\pi$.

We saw earlier that the M-theory origin of the D-2-brane requires $iV/2\pi$ to be
a $U(1)$ gauge potential. It is then a consequence of T-duality that $iV/2\pi$
is equally a $U(1)$ gauge potential for any D-brane, in particular the D-string.
Because of this, the 1-form $V$ of the D-string action is defined only up to the
$U(1)$ gauge transformation
\begin{equation}
{iV\over 2\pi} \rightarrow {iV\over 2\pi} + g^{-1}dg \qquad
\big(g(t,\sigma) \in U(1)\big)\ .
\label{eq:gtran}
\end{equation}
We may choose $g= e^{i\sigma}$, in which case the gauge transformation becomes
\begin{equation}
V_\sigma\rightarrow V_\sigma + 2\pi\ . 
\label{eq:nonconta}
\end{equation} 
Since this is a {\it gauge} transformation, we must identify $V_\sigma$ with
its gauge transform $V_\sigma + 2\pi$. Thus $V_\sigma$ is the coordinate of a
compact direction with the standard identification, so $2\pi R_A$ is the length of
the closed orbit of $\tilde k=\partial/\partial V_\sigma$, i.e. $R_A$ is the
radius of the compact dimension, as required.

We have now established the relation of the IIB D-string to the IIA D-0-brane.
It is similarly related to the D-2-brane. Let us now investigate its relation
to the IIB Fundamental string, or `F-string'. Since $V_\sigma$ in
(\ref{eq:hamstring}) is identified with period $2\pi$, the eigenvalues of its
conjugate variable $E$ are integers. Let us choose 
\begin{equation}
E=m\, ;
\label{tendim}
\end{equation}
the $V_t E'$ term is then zero and the $\dot V_\sigma E$ term becomes a total
derivative which may be neglected. The Lagrangian density of
the action (\ref{eq:hamstring}) is thereby reduced to
\begin{equation}
{\cal L} = \dot X\cdot P + s\; X'\cdot P  -{1\over2}v {\cal H}
\label{tendima}
\end{equation}
where
\begin{equation}
{\cal H}= (P - n{\cal B}' +m{\cal B})^2
+ (X')^2[ (m-n\ell)^2 +n^2e^{-2\phi_B}] 
\label{eq:tendimc}
\end{equation}
is the `Hamiltonian' constrained to vanish by the Lagrange multiplier $v$.
Setting the background scalar fields to their vacuum values, we see that
this is the action for a string with tension \cite{schm,alwis}
\begin{equation}
T= {1\over 2\pi}\sqrt{ n^2/g_s^2 + (m-n\langle\ell\rangle )^2}
\label{eq:tendimd}
\end{equation}
and charge $(m,n)$ with respect to $(B,-B')$, exactly as required
\cite{schwb,witbrane} by the $Sl(2;Z)$ symmetry of the IIB theory. In
particular, {\it the string with charge $(1,0)$ is just the Fundamental IIB
string}. 

It would make no sense to set $n=0$ in the original D-string action
(\ref{eq:dstring}), but this is a defect of that action rather than a physical
limitation. One of the virtues of the Hamiltonian form of the action is
that it makes this fact manifest. Indeed, setting $n=0$ in (\ref{eq:tendimc}) and
eliminating all auxiliary variables from the action one recovers, provided that
$m\ne0$, the fundamental string action (\ref{eq:dimrede}) with $\nu=m$, i.e.
with tension $T=m/2\pi$; that action was derived from M-theory as the action of
the IIA string but since we are omitting fermions it is equally the bosonic
sector of the action for the IIB string. Thus, the fundamental string tension can
be identified as the lowest non-zero eigenvalue of $E$. The semi-classical
equivalent of an eigenvalue of $E$ is the circulation of the classical variable
$E(\sigma)$ around the string, ${1\over2\pi}\int d\sigma E(\sigma)$. This
equals the flux of the BI 2-form through the string worldsheet. Thus, up to a
normalization factor, {\it the fundamental string tension is the quantized flux of
the BI 2-form through the string worldsheet} \cite{london}. 

No further details of IIB p-branes will be given here, but some mention must be
made of the IIB self-dual D-3-brane. In many respects, this plays as crucial a
role in the IIB theory as the D-2-brane does in the IIA theory. The effective
action for the D-3-brane is of the form (\ref{eq:bosmemh}). An important
feature of its equations of motion (the `branewave' equations) is that
they exhibit an $Sl(2;Z)$ `duality' in the sense that an $Sl(2;Z)$
transformation of the worldvolume fields, which acts by a generalization of
electromagnetic duality on the BI 2-form field strength and its Hodge dual, 
effects an $Sl(2;Z)$ transformation of the IIB supergravity
background \cite{tsey}. Thus, the $Sl(2;Z)$ invariance of IIB supergravity
(actually $Sl(2;R)$ but this is broken to $Sl(2;Z)$ in the quantum superstring
theory) extends to the combined supergravity plus branewave equations. 
Given that M-theory predicts both the $Sl(2;Z)$ symmetry (as the modular group
of a 2-torus) {\it and} the 3-brane (as a $T^2$-wrapped 5-brane), this result is
clearly a consequence of M-theory. From this perspective it is also clear that
M-theory equally predicts an $Sl(2;Z)$ duality `on the brane' for $n$ coincident
3-branes, for which the BI $U(1)$ group is enhanced to $U(n)$. After
gauge-fixing the $\kappa$-symmetry and ignoring all but the leading order
terms in an $\alpha'$ expansion, the worldvolume field theory of this
multi-3-brane is just an N=4 D=4 super-YM theory, for which we can interpret the
predicted $Sl(2;Z)$ duality as the conjectured S-duality of this
theory \cite{font,sens}.  
 
Let us also call the $Sl(2;Z)$ duality of IIB superstring theory `S-duality',
since in both the D=10 and D=4 contexts there is a $Z_2$ subgroup that 
interchanges weak and strong coupling. As just explained, S-duality of IIB
superstring theory can be `derived' from the electromagnetic S-duality of the
3-brane in essentially the same way that spacetime T-duality is derived from 
duality on the worldsheet of the fundamental string. In the former case 
duality `on the brane' exchanges a D=4 vector potential with its
electromagnetic dual vector potential whereas in the latter case it exchanges a
scalar for its dual scalar. In both cases, a duality transformation `on the
brane' results in a duality transformation of the background spacetime fields.
We have also seen in this lecture how a vector to scalar duality on the
worldvolume of the D-2-brane results in a transformation from the background
fields of D=10 IIA supergravity to those of D=11 supergravity. Let us call the
latter transformation `M-duality'. Then, as illustrated in Table 2, {\it all} the
Type II dualities can be seen to have a common origin in dualities `on the brane'.
\begin{table}[t]
\caption{{\bf spacetime dualities from dualities `on the brane'}}
\begin{center}
\begin{tabular}{||c|c|c||}
\hline
&Worldvolume duality & Spacetime duality \\
\hline\hline
T & p=1 : $\phi\leftrightarrow\tilde\phi$ & $IIA\leftrightarrow IIB$ \\
\hline
M & p=2 : $V\leftrightarrow \phi$ & $IIA \leftrightarrow M$ \\
\hline
S & p=3 : $V\leftrightarrow \tilde V$ & $IIB\leftrightarrow IIB$ \\
\hline\hline
\end{tabular}
\end{center}
\end{table}
It  has been argued \cite{senrev} that there is only one other M-theory or
superstring duality that is `independent' of these Type II dualities, and that it
can be taken to be the Type I to $SO(32)$ heterotic string duality. Using these
four dualities one can get to any brane on the M-theory brane scan from any
other one. In other words, M-theory as we now know it is a `p-brane democracy'.

\section{Epilog}

The main aim of these lectures has been to explain how the five D=10 superstring
theories are unified by 11-dimensional M-theory. Pedagogical expediency has
dictated the omission of many other interesting topics, in particular connections
between superstring and M-theory compactifications in lower dimensions. Perhaps
the gravest omission is a definition of M-theory. One excuse for this is that
whereas definitions may come first in mathematics they usually come last in
physics. It therefore seems appropriate to end these lectures with a brief
mention of recent progress on this front. The obvious starting point for a
definition of M-theory is the D=11 supermembrane. In the past, there were two
major objections to a fundamental supermembrane theory. These were (i) that the
(2+1)-dimensional worldvolume action is non-renormalizable and (ii) that the
spectrum of the first quantized supermembrane is continuous. Both these problems
now have answers. 

The non-renormalizability problem has been overcome by an interpretation of both
the (1+1)-dimensional D=10 superstring actions, and the (2+1)-dimensional D=11
supermembrane action as effective actions of the so-called `n=2 heterotic
strings'. In this approach \cite{mart}, the classical supermembrane equations
emerge as the conditions required for conformal invariance of the n=2 heterotic
string sigma-model action, so the first quantized supermembrane is interpreted as
a second-quantized string theory. The second quantized supermembrane would
presumably then emerge from a `third-quantized n=2 heterotic string theory'.
Since we have little idea what this might be, the non-renormalizability problem
might appear to have been solved only at the cost of introducing a new problem.
On the other hand, there are some indications from an alternative approach
\cite{BFSS} that first quantization of the supermembrane (and hence
second-quantization of the n=2 heterotic string) might be sufficient. 

This alternative approach makes use of the observation \cite{pktmbrane} that the
large N matrix model approximation to the supermembrane Hamiltonian
\cite{nicolai} can be re-interpreted as the Hamiltonian for N coincident
D-0-branes; the continuity of the spectrum is then seen to be a consequence
of the no-force condition between D-0-branes. This suggests a
re-interpretation of the Hilbert space of the first quantized supermembrane as
the Hilbert space of an interacting multi-particle (and multi-membrane) theory
that one would normally expect to arise only on second quantization. Remarkably,
it seems possible to extract sensible results for the scattering of D=11
gravitons, and to recover both the membrane and the fivebrane as collective
excitations in this aproach \cite{BFSS,dougfive}. Is this the long sought theory
of quantum gravity? If past experience is anything to go by, the future holds
plenty of surprises in store for us.

\end{document}